\documentclass[
nofootinbib,
amsmath,amssymb,
prb,
twocolumn,
]{revtex4-2}
\usepackage{dcolumn}
\usepackage{bm}
\usepackage{braket}
\usepackage{color}
\usepackage{accents}
\usepackage{mathrsfs}
\usepackage{bm}
\usepackage{here}
\usepackage{longtable}

\usepackage{mathtools}
\usepackage{empheq}
\usepackage{graphicx}
\usepackage{hyperref}
\usepackage{comment}
\usepackage{float}

\begin{document}
\title{
Quantum theory of magnetic octupole in periodic crystals and 
application to $d$-wave altermagnets
}
\author{Takumi Sato$^{\ast}$}
\author{Satoru Hayami$^{\dag}$}
\affiliation{
Graduate School of Science, Hokkaido University, Sapporo 060-0810, Japan\\
\textnormal{$^{\ast}$sato@phys.sci.hokudai.ac.jp\\
\textnormal{$^{\dag}$hayami@phys.sci.hokudai.ac.jp}}
}

\date{\today}

\begin{abstract}
\noindent\textbf{Abstract}\quad
Magnetic multipoles have been recognized as order parameters characterizing magnetic structure in solids. 
Recently, magnetic octupoles have been proposed as the order parameters of time-reversal-symmetry breaking centrosymmetric antiferromagnets exhibiting nonrelativistic spin splitting, which is referred to as ``altermagnet''. 
However, a gauge-invariant formulation of magnetic octupoles in crystalline solids remains elusive. 
Here, we present a gauge-invariant expression of spin magnetic octupoles in periodic crystals based on quantum mechanics and thermodynamics, which can be used to quantitatively characterize time-reversal-symmetry breaking antiferromagnets including $d$-wave altermagnets. 
The allowed physical response tensors are classified beyond symmetry considerations, and direct relationships are established for some of them in insulators at zero temperature. 
Furthermore, our expression reveals a contribution from an anisotropic magnetic dipole, which has the same symmetry as conventional spin and orbital magnetic dipoles but carries no net magnetization. 
We discuss the relation between the anisotropic magnetic dipole and the anomalous Hall effect.
\end{abstract}
\maketitle

\section*{Introduction}
Magnetism is a fundamental form of order in solids, long studied for its rich physics and functional significance, driven primarily by the electron spin degree of freedom and its interactions.
Owing to both its rich physical origins and technological relevance, magnetic order continues to be a central topic in condensed matter physics. 
A wide variety of magnetic structures emerge depending on the nature of electron correlations.
Therefore, identifying suitable order parameters is essential for understanding and controlling these phases.
In conventional ferromagnets (FMs) such as Fe, Co, and Ni, the net spin magnetization serves as the primary order parameter,
which is defined as
\begin{align}
    M_i \coloneqq
    g\mu_{\mathrm{B}} \sum_{n} \int \frac{d^{d}k}{(2\pi)^{d}} \Braket{\psi_{n\bm{k}}|\hat{s}_i|\psi_{n\bm{k}}} f_{n\bm{k}} ,
    \label{eq:MD}
\end{align}
where $i=x,y,z$; $g$, $\mu_{\mathrm{B}}$, $d$,
$\Ket{\psi_{n\bm{k}}}$, $\hat{\bm{s}}$, and $f_{n\bm{k}}$ represent
spin g-factor, Bohr magneton, spatial dimension,
the Bloch state, spin operator, and Fermi distribution function, respectively.
Here, we define the ``order parameter'' as a tensor characterizing the (magnetic) order 
of the condensate and encoding the electronic structure. 
As seen from Eq.~\eqref{eq:MD}, it involves the Bloch wave functions and Fermi distribution functions 
and can be expressed as the expectation value of a multipole moment. 
Although it shares the same symmetry properties as the order parameters in Landau theory, 
it is essentially different (as will be discussed in Discussion section).
The magnetization transforms as a time-reversal-odd rank-1 axial tensor and hence regarded as a magnetic dipole (MD).
This makes ferromagnetism directly responsive to external magnetic fields,
which are the conjugate fields of the MD.

On the contrary, the identification of an appropriate order parameter for antiferromagnets (AFMs) remains nontrivial 
because they have no net magnetization and do not couple directly to uniform magnetic fields.
Although N\'{e}el vector~\cite{neel:annphys1936}, which is defined as the difference
in local magnetic moments between sublattices, has long been used as a quantity that characterizes collinear AFMs,
it is inherently a local quantity without a well-defined conjugate field in the thermodynamics~\cite{bhowal:prx2024}
and cannot describe magnetic structures without any ambiguity.
Moreover, recent studies have unveiled an unconventional
class of collinear AFMs that host
nonrelativistic momentum-dependent spin splitting in the band structures when time-reversal symmetry is broken%
~\cite{noda:pccp2016,okugawa:jpcm2018,smejkal:sciadv2020,naka:natcomm2019,ahn:prb2019,
hayami:jpscp2020,hayami:jpsj2019spinsplit,hayami:prb2020spinsplit,yuan:prb2020,naka:prb2021perovskite,
yuan:prm2021,yuan:prb2021,hernandez:prl2021,smejkal:prx2022magnetoresistance,mazin:pnas2021,
cheong:npj2025AMclassification,guo:advmat2025review,hu:prx2025}.
Such AFMs have recently come to be referred to as altermagnets~\cite{smejkal:prx2022spingroup1,smejkal:prx2022spingroup2}.
In altermagnets, spin sublattices in the unit cell are related by a rotation of the crystal.
Theoretical and experimental studies have demonstrated that physical phenomena traditionally associated with FMs
such as the anomalous Hall effect (AHE) can also emerge
in such AFMs~\cite{solovyev:prb1997,smejkal:sciadv2020,shao:prapp2021,samanta:jappphys2020,sivadas:prl2016,
naka:prb2020,naka:prb2021AHEperovskite,attias:prb2024,chen:prb2022,smejkal:natrevmat2022}.
This has motivated the study for a clear and
physically grounded definition of an appropriate order parameter characterizing AFMs without time-reversal symmetry.

To resolve the issue, a higher-rank magnetic
multipole~\cite{kimura:npj2021,jeon:npj2024,okamoto:npj2025} has been proposed to characterize centrosymmetric AFMs~\cite{suzuki:prb2017,bhowal:prx2024}.
The coupling between the local magnetic moment $\bm{\mu}(\bm{r})$
and a magnetic field $\bm{B}(\bm{r})$ in free-energy can be expressed as
\begin{align}
    F
    &= - \int d^{d} r \, \mu_i(\bm{r}) \, B_i(\bm{r}) \nonumber \\
    &= - \int d^{d} r \, \mu_i(\bm{r}) B_i(\bm{0})
       - \int d^{d} r \, \mu_i(\bm{r}) r_j \partial_{r_j} B_i(\bm{0}) \nonumber \\
    &\hspace{1em} - \frac{1}{2} \int d^{d} r \, \mu_i(\bm{r}) r_j r_k \,
    \partial_{r_j} \partial_{r_k} B_i(\bm{0}) - \cdots ,
    \label{eq:free_energy_expansion}
\end{align}
where $B_i(\bm{r})$ is expanded around $\bm{r}=\bm{0}$ to reach the second and third lines~\cite{spaldin:jpcm2008,bhowal:prx2024}.
Zeroth and first-order terms in Eq.~\eqref{eq:free_energy_expansion} vanish in centrosymmetric AFMs,
leaving the third-order term as the leading contribution.
This suggests the magnetic octupole (MO) $\int d^{d} r \, \mu_i(\bm{r}) r_j r_k$,
as a natural order parameter for such AFMs, conjugate to the
spatial variation of a magnetic field: $\partial_{r_j} \partial_{r_k} B_i$.

Formally, the spin MO in crystalline solids seems to be defined as
\begin{align}
    M_{ijk} \overset{?}{\coloneqq}
    g\mu_{\mathrm{B}} \sum_{n} \int \frac{d^{d}k}{(2\pi)^{d}}
    \Braket{\psi_{n\bm{k}}|\hat{s}_i \hat{r}_j \hat{r}_k|\psi_{n\bm{k}}} f_{n\bm{k}} ,
    \label{eq:MOM}
\end{align}
with $\hat{\bm{r}}$ being the position operator.
Unfortunately, this expression is not gauge invariant since the position operator is ill-defined in the Bloch basis.
This kind of difficulty in evaluating multipoles in solids occurs frequently and theoretical efforts have solved this difficulty.
For example, gauge-invariant formulations in terms of Bloch functions have been presented for 
electric dipole (ED), i.e., electric polarization, which is, strictly speaking, uniquely determined modulo the lattice points%
~\cite{king-smith:prb1993, vanderbilt:prb1993, resta:rmp1994, resta:book, resta:jpcm2010review, xiao:rmp2010, vanderbilt:book}; 
orbital MD (orbital magnetization)%
~\cite{resta:chemphyschem2005, xiao:prl2005, xiao:prl2006thermoele, thonhauser:prl2005, ceresoli:prb2006,
shi:prl2007, souza:prb2008, xiao:rmp2010, thonhauser:ijmpb2011, vanderbilt:book}; 
spin and orbital magnetic quadrupoles (MQs)%
~\cite{gao:prb2018spinMQM, shitade:prb2018orbitalMQM,gao:prb2018orbitalMQM, shitade:prb2019spinMQM}; 
and electric quadrupole (EQ)~\cite{daido:prb2020}.
However, the gauge-invariant expression of the MO has remained missing.

In this work, we address the issue on spin MO
based on quantum mechanics and thermodynamics%
~\cite{shi:prl2007, gao:prb2018spinMQM, shitade:prb2018orbitalMQM,
gao:prb2018orbitalMQM, shitade:prb2019spinMQM, daido:prb2020}
and obtain a gauge-invariant expression.
We find that some of the physical response tensors, expected in 
MO orderings from spacetime symmetries and tensor ranks, 
are directly related to the gauge-invariant MO through a thermodynamic relation. 
We also classify physical responses expected in not only the MO ordering but also MD and MQ orderings
from a thermodynamic viewpoint that goes beyond symmetry arguments.
Through model calculations, we demonstrate
that the nonzero components of the gauge-invariant $M_{ijk}$ appropriately capture the magnetic structures.
In addition, we show that one of the dipole components obtained by the irreducible decomposition of $M_{ijk}$
corresponds to the anisotropic magnetic dipole (AMD),
a time-reversal-odd axial vector that does not carry net magnetization~\cite{kusunose:jpsj2020}.
Notably, we find that the gauge-invariant AMD faithfully captures the magnetic structure of AFMs exhibiting
AHE~\cite{hayami:prb2021AMD} from symmetry considerations.
Our results establish the gauge-invariant $M_{ijk}$,
as physically meaningful descriptors of AFMs without time-reversal symmetry,
which will be useful to characterize the nature of altermagnets from the microscopic viewpoint.
This is useful to characterize the magnetic structure within a unified
framework and allows for quantitative analysis, while the N\'{e}el vector does not.

The rest of this paper is organized as follows.
In Results section, we present the gauge-invariant formulation of spin MO
based on quantum mechanics and thermodynamics.
We also discuss the relationships between physical response tensors and the MO based on thermodynamics.
We also present the results of the model calculation
to analyze the behavior of the MO in a collinear AFM and FM.
In Discussion section, we discuss the obtained results,
including the relationship between the AMD and the AHE.
Throughout this paper, we use the units of $k_\mathrm{B} = c = \hbar = 1$
where $k_\mathrm{B}$ is the Boltzmann constant and $c$ is the speed of light
and $e<0$ is charge of an electron.

\section*{Results}
\subsection*{Spin MO in crystalline solids}
We begin with the differential form of the free-energy density $F(\bm{r})$ in crystalline solids with a slowly varying magnetic field
$\bm{B}(\bm{r})$~\cite{shi:prl2007,gao:prb2018spinMQM,shitade:prb2018orbitalMQM,gao:prb2018orbitalMQM,shitade:prb2019spinMQM}:
\begin{align}
    dF =
    &-SdT - Nd\mu - M_{i} dB_{i} \nonumber \\
    &-M_{ij} d[\partial_{r_j} B_{i}] - M_{ijk} d[\partial_{r_j} \partial_{r_k} B_{i}] ,
    \label{eq:differential_FE}
\end{align}
where $S$, $T$, $N$, and $\mu$ are the entropy, temperature, particle number, and chemical potential, respectively;
$M_i$, $M_{ij}$, and $M_{ijk}$ for $i,j,k = x, y, z$
are the spin MD (spin magnetization), spin MQ, and spin MO, respectively.
Here, we focus on the contributions from the magnetic multipoles originating from spin 
and do not discuss the orbital part. The latter is discussed in our recent work~\cite{TS:arxiv2025orbitalMO}.
Nevertheless, the discussion of the thermodynamic relation to the response tensors is also valid for the orbital part 
(see the next subsection for details).
We define the spin MO in crystalline solids as
\begin{align}
    M_{ijk} \coloneqq - \left( \frac{\partial F}{\partial[\partial_{r_j} \partial_{r_k} B_i]} \right)_{T,\mu,\bm{B},\partial_{\bm{r}}\bm{B}} .
    \label{eq:def_MOM}
\end{align}
For a practical calculation, it is convenient to define the auxiliary quantity:
\begin{align}
    \tilde{M}_{ijk} \coloneqq - \left( \frac{\partial K}{\partial[\partial_{r_j} \partial_{r_k} B_i]} \right)_{T,\mu,\bm{B},\partial_{\bm{r}}\bm{B}} ,
    \label{eq:def_tildeMOM}
\end{align}
with $K \coloneqq F+TS$ being the energy density.
Using the Maxwell relation,
the desired spin MO is obtained by solving the differential equation~\cite{shi:prl2007}:
\begin{align}
    \frac{\partial (\beta M_{ijk})}{\partial \beta} = \tilde{M}_{ijk} ,
    \label{eq:differential}
\end{align}
where $\beta = 1/T$.

To obtain $\tilde{M}_{ijk}$, we make use of the Kubo formula~\cite{shitade:prb2019magnonGME,daido:prb2020}.
We calculate the energy-magnetization correlation function,
\begin{align}
    \chi_{K, M_i} &(\bm{q},\omega) = \nonumber \\
    &-g\mu_{\mathrm{B}}\sum_{nm} \int \frac{d^d k}{(2\pi)^d} \frac{\epsilon_{n\bm{k}} + \epsilon_{m\bm{k}+\bm{q}}-2\mu}{2} \nonumber \\
    &\times \Braket{u_{n\bm{k}}|u_{m\bm{k}+\bm{q}}} \Braket{u_{m\bm{k}+\bm{q}}|\hat{s}_{i}|u_{n\bm{k}}} \nonumber \\
    &\times \frac{f(\epsilon_{n\bm{k}}-\mu)-f(\epsilon_{m\bm{k}+\bm{q}}-\mu)}{\omega+i\delta + \epsilon_{n\bm{k}} -\epsilon_{m\bm{k}+\bm{q}}} .
\end{align}
Here, we used the following notations:
\begin{align}
    \hat{H} \Ket{\psi_{n\bm{k}}} &= \epsilon_{n\bm{k}} \Ket{\psi_{n\bm{k}}} , \quad
    \Ket{u_{n\bm{k}}} = e^{-i\bm{k}\cdot \hat{\bm{r}}} \Ket{\psi_{n\bm{k}}} , \\
    f(z) &= [e^{\beta z}+1]^{-1} ,
    \label{eq:chi}
\end{align}
where $\hat{H}$ is the single-electron Hamiltonian in a periodic crystal.
The change of the energy is examined by
$\delta \langle K \rangle (\bm{q},\omega) = \chi_{K, M_i} (\bm{q},\omega) B_i (\bm{q},\omega)$ and
we define the auxiliary MO as
$\tilde{M}_{ijk} = \lim_{\bm{q} \rightarrow \bm{0}}
\partial_{q_j}\partial_{q_k} \lim_{\delta \rightarrow +0} \chi_{K, M_i} (\bm{q},0)/ 2$.
By solving Eq.~\eqref{eq:differential}, we obtain the spin MO
in crystalline solids:
\begin{widetext}
\begin{align}
    M_{ijk} &= g \mu_{\mathrm{B}} \sum_{n} \int \frac{d^d k}{(2\pi)^d}
    \left( S_{n,\text{diag}}^{ijk} + S_{n,\text{off-diag}}^{ijk} \right)
    \label{eq:Mijk} , \\
    S_{n,\text{diag}}^{ijk} &=
    -\frac{1}{12} s^i_n \partial_{k_j} \partial_{k_k} \epsilon_n f_n'
    +\sum_{m}^{\neq n} \left[
        \frac{1}{2} (s^i_n+s^i_m) \left\{ g^{jk}_{nm} f_n + G^{jk}_{nm} \int_{\epsilon_n-\mu}^{\infty} dz f(z) \right\} 
    \right] \label{eq:S_diag} , \\
    S_{n,\text{off-diag}}^{ijk} &=
    \Biggl(
    \sum_{m}^{\neq n} \left[
        \frac{1}{6} \tilde{d}^{ij}_{nm} v^k_n f_n' + \frac{1}{2}\tilde{g}^{ij}_{nm} v^k_n f_n
        + \frac{1}{4} \tilde{G}^{ij}_{nm} ( v^k_n + v^k_m ) \int_{\epsilon_n-\mu}^{\infty} dz f(z)
        \right] \nonumber \\
    &\hspace{1em}+\sum_{m}^{\neq n} \sum_{\ell}^{\neq n,m} \left[
        -\frac{1}{4} X_{nm\ell}^{ijk} f_n -\frac{1}{2} Y_{nm\ell}^{ijk} \int_{\epsilon_n-\mu}^{\infty} dz f(z)
        \right] \Biggr) + ( j \leftrightarrow k ) ,
        \label{eq:S_off-diag}
\end{align}
\end{widetext}
where
\begin{align}
    g^{ij}_{nm} &=  \mathrm{Re} \left[ \frac{v^i_{nm} v^j_{mn}}{\epsilon_{nm}^2} \right] , \ 
    G^{ij}_{nm}  = 2\mathrm{Re} \left[ \frac{v^i_{nm} v^j_{mn}}{\epsilon_{nm}^3} \right] 
    \label{eq:g} , \\
    \tilde{g}^{ij}_{nm} &=  -\mathrm{Re} \left[ \frac{s^i_{nm} v^j_{mn}}{\epsilon_{nm}^2} \right] , \ 
    \tilde{G}^{ij}_{nm} = -2\mathrm{Re} \left[ \frac{s^i_{nm} v^j_{mn}}{\epsilon_{nm}^3} \right]
    \label{eq:tilg1} , \\
    \tilde{d}^{ij}_{nm} &= \mathrm{Re} \left[ \frac{s^i_{nm} v^j_{mn}}{\epsilon_{nm}} \right] 
    \label{eq:tilg2} , \\
    X_{nm\ell}^{ijk} &= \mathrm{Re} \left[
    \frac{s^i_{m \ell}v^j_{\ell n} v^k_{nm}}{\epsilon_{\ell n} \epsilon_{nm}} \right]
    \label{eq:X} , \\
    Y_{nm\ell}^{ijk} &= \nonumber \\
    &\hspace{-2em}\mathrm{Re} \left[
        -\frac{s^i_{nm}v^j_{m\ell} v^k_{\ell n}}{\epsilon_{nm} \epsilon_{m\ell} \epsilon_{\ell n}}
        + \frac{s^i_{\ell n}v^j_{nm} v^k_{m\ell}}{\epsilon_{nm}^2 \epsilon_{m\ell}}
        + \frac{s^i_{m \ell}v^j_{\ell n} v^k_{nm}}{\epsilon_{\ell n} \epsilon_{nm}^2}
    \right] \label{eq:Y} .
\end{align}
The following abbreviations were used in Eqs.~\eqref{eq:S_diag}-\eqref{eq:Y}:
$s^i_{nm} = \Braket{u_{n\bm{k}}|\hat{s}_i|u_{m\bm{k}}}$, $s^i_{n} = s^i_{nn}$,
$v^i_{nm} = \Braket{u_{n\bm{k}}|\partial_{k_i} \hat{H}_{\bm{k}}|u_{m\bm{k}}}$, $v^i_{n} = v^i_{nn}$,
$\epsilon_n = \epsilon_{n\bm{k}}$, $\epsilon_{nm} = \epsilon_n - \epsilon_m$,
$f_n = f(\epsilon_{n}-\mu)$, and $f_n' = {\partial f(z) / \partial z}|_{z=\epsilon_{n}-\mu}$ 
with $\hat{H}_{\bm{k}} = e^{-i\bm{k}\cdot \hat{\bm{r}}} \hat{H} e^{i\bm{k}\cdot \hat{\bm{r}}}$ being the Bloch Hamiltonian.
The symbol $+ (j \leftrightarrow k)$ in Eq.~\eqref{eq:S_off-diag} denotes the addition of a term obtained by exchanging the indices $j$ and $k$
in the preceding term: $A^{ijk} + \left( B^{ijk} + (j \leftrightarrow k)\right)= A^{ijk} + B^{ijk} + B^{ikj}$.
Equation~\eqref{eq:Mijk} is invariant under the gauge transformation:
$\ket{u_{n\bm{k}}} \rightarrow e^{i\eta_{n}(\bm{k})} \ket{u_{n\bm{k}}}$ ($\eta_{n}(\bm{k}) \in \mathbb{R}$).
The expression in the presence of the degeneracy points is presented in Methods section.

Equation~\eqref{eq:Mijk} contains the quantities which characterize the geometry
of the eigenstate space of the parameter-dependent Hamiltonian:
$\hat{H}_{\bm{k},\bm{h}} = \hat{H}_{\bm{k}} -  \hat{\bm{s}} \cdot \bm{h}$ with $\bm{h} = g\mu_{\mathrm{B}} \bm{B}$.
Hereafter, we refer to the three-dimensional parameter space 
spanned by the momentum ($k_x, k_y, k_z$) as $k$-space, and
six-dimensional parameter space spanned by the momentum and the Zeeman field 
($k_x, k_y, k_z, h_x, h_y, h_z$) as $kh$-space.
A rank-$\ell$ polar (axial) tensor is defined as a tensor whose spatial parity is $(-1)^{\ell}$ ($(-1)^{\ell+1}$).
In the spin-diagonal component $S_{n,\text{diag}}^{ijk}$, 
$g^{ij}_{n} = \sum_{m}^{\neq n}g^{ij}_{nm}$ corresponds to the $k$-space quantum metric (QM),
which is the real part of the quantum geometric tensor (QGT) (or Fubini-Study metric)
$T^{ij}_n = \Braket{\partial_{k_i} u_{n\bm{k}}|
(1-\Ket{u_{n\bm{k}}}\Bra{u_{n\bm{k}}})|\partial_{k_j} u_{n\bm{k}}}$~\cite{provost:cmp1980,berry:book1989,resta:epjb2011}.
$G^{ij}_{n} = \sum_{m}^{\neq n}G^{ij}_{nm}$ corresponds to the $k$-space Berry connection polarizability (BCP) (or positional shift),
which corresponds to the first-order correction to the Berry connection by an electric field~\cite{gao:prl2014BCP}.
$k$-space QM, BCP, and $\partial_{k_i} \partial_{k_j} \epsilon_n$ are transformed as time-reversal-even rank-2 polar tensors.
Therefore, based on their symmetry and rank, they can be regarded as EQs.
Consequently, the spin-diagonal component $S_{n,\text{diag}}^{ijk}$ in Eq.~\eqref{eq:Mijk}
can be regarded as the tensor products of the MDs and the EQs.
Actually, replacing $g\mu_{\mathrm{B}}\hat{\bm{s}}$ by $1$ in Eq.~\eqref{eq:Mijk}, 
we obtain the thermodynamic EQ~\cite{daido:prb2020}.

In the spin-off-diagonal component $S_{n,\text{off-diag}}^{ijk}$,
$\tilde{g}^{ij}_{nm}$ and $\tilde{G}^{ij}_{nm}$ are related to the $kh$-space QM and BCP
($\tilde{g}^{ij}_{n}=\sum_{m}^{\neq n}\tilde{g}^{ij}_{nm}$ and
$\tilde{G}^{ij}_{n}=\sum_{m}^{\neq n}\tilde{G}^{ij}_{nm}$)~\cite{xiao:prl2022hkspaceBCP,feng:mtq2025}.
In addition, $\tilde{d}^{ij}_{n}=\sum_{m}^{\neq n}\tilde{d}^{ij}_{nm}$
is related to the real part of the geometric quantity in the $kh$-space:
$\Bra{\partial_{h_i} u_{n\bm{k},\bm{h}}} \left( \hat{H}_{\bm{k},\bm{h}} - \epsilon_{n\bm{k},\bm{h}} \right)
\Ket{\partial_{k_j} u_{n\bm{k},\bm{h}}}$%
~\cite{resta:arxiv2017DWOAM,resta:jpcm2018DWSCW,kang:natphys2024}.
These three geometric terms are transformed as time-reversal-even rank-2 axial tensors,
and hence correspond to the electric toroidal quadrupoles as determined by their spacetime symmetry and rank.
Therefore, parts of $S_{n,\text{off-diag}}^{ijk}$ consist of the tensor products of the electric toroidal 
quadrupole and the magnetic toroidal dipoles, which is time-reversal-odd rank-1 polar tensor
(group velocity $v^k_{n}$)~\cite{hayami:jpsj2024review}.
The rest of $S_{n,\text{off-diag}}^{ijk}$ is composed of
the tensor products of an MD and two magnetic toroidal dipoles, 
which satisfies the symmetry properties of MO
and may be associated with higher-order geometric quantities~\cite{hetenyi:pra2023}.

There is a structural difference between the spin MO and the spin MQ%
~\cite{gao:prb2018spinMQM,shitade:prb2019spinMQM}.
The spin MO involves only the real parts of geometric quantities such as QM, 
whereas the spin MQ involves only the imaginary parts of geometric quantities such as the Berry curvature 
in $kh$-space~\cite{gao:prb2018spinMQM,shitade:prb2019spinMQM}.
This distinction arises because a change in the tensor rank by one leads to an interchange 
between real and imaginary parts, as the position operator in momentum space is given by 
$\bm{r} \sim i\nabla_{\bm{k}}$. 
This also explains the structural similarities
between the spin MQ and orbital MD
as the orbital MD is obtained from the antisymmetric part of orbital magnetic toroidal quadrupole (MTQ)
$T^{\mathrm{orb}}_{ij} \sim \int d \bm{r} \, r_i v_j$%
~\cite{gao:prb2018spinMQM,shi:prl2007}.

$M_{ijk}$ is a time-reversal-odd rank-3 axial tensor that is symmetric
under the exchange of the second and third indices ($j$ and $k$)~\cite{bhowal:prx2024,oike:prb2024}.
It has 18 components and can be decomposed into
(totally symmetric) MO ($M_{3m}$),
MTQ ($T_{2m}$), and two MDs ($M_{1m}$ and $M'_{1m}$)
satisfying the decomposition of an 18-dimensional reducible representation
of SO(3): $18 = 7 \oplus 5 \oplus 3 \oplus 3$
(for example, see Ref.~\cite{urru:annphys2022}).
We decompose $M_{ijk}$ as
\begin{align}
    M_{ijk} &=
    M_{ijk}^{M_{3m}} + M_{ijk}^{T_{2m}} + M_{ijk}^{M_{1m}} + M_{ijk}^{M'_{1m}} 
    \label{eq:MO_decomposition},
\end{align}
where each term represents, in order, the contributions from the MO, MTQ, MD, and the other MD.
The expressions of the decomposed tensors are presented in Supplemental Material~\cite{supple}.
In general, all terms in Eq.~\eqref{eq:Mijk} can contribute to each decomposed tensor, because
they share the same symmetry properties.
Notably, we find one of the MDs ($M'_{1m}$) corresponds to anisotropic magnetic dipole (AMD) 
defined by $M'_{i} \coloneqq \left[3M_{jji}-M_{ijj}\right]/\sqrt{10}$~\cite{kusunose:jpsj2020} 
with the unit of $\mu_{\mathrm{B}}$, using the lattice constant as the length unit.
Originally proposed as part of a complete basis set for electronic degrees of freedom 
in isolated atoms~\cite{kusunose:jpsj2020}, 
AMD has the same symmetry properties as the ordinary MDs such as spin and orbital angular momenta
but does not carry a net magnetic moment.
It should be noted that our decomposition of the dipole components differs from that in Ref.~\cite{urru:annphys2022}; 
however, the total dipole components ($M_{ijk}^{M_{1m}} + M_{ijk}^{M'_{1m}}$)
span the same subspace as in Ref.~\cite{urru:annphys2022}.
The properties of the decomposed tensors will be discussed in the next section.

\subsection*{Thermodynamic relations and response tensors}
The MO relates to physical response tensors.
To provide a comprehensive discussion of the thermodynamic magnetic multipoles, 
we consider the orbital MD and both the spin and orbital contributions to the MQ and MO here.
The Maxwell relations
\begin{align}
    \frac{\partial M_{i}}{\partial \mu}   &= \frac{\partial N}{\partial B_i} 
    \label{eq:maxwell1}, \\
    \frac{\partial M_{ij}}{\partial \mu}  &= \frac{\partial N}{\partial[\partial_{r_j} B_i]} 
    \label{eq:maxwell2}, \\
    \frac{\partial M_{ijk}}{\partial \mu} &= \frac{\partial N}{\partial[\partial_{r_j} \partial_{r_k} B_i]}
    \label{eq:maxwell3},
\end{align}
hold true from Eq.~\eqref{eq:differential_FE}.
For later use, we define the derivatives of the magnetic multipoles
with respect to the chemical potential as
\begin{align}
    \alpha_i &\coloneqq e \frac{\partial M_{i}}{\partial \mu} \label{eq:_MD},\\
    \alpha_{ij}  &\coloneqq -e   \frac{\partial M_{ij}}{\partial \mu} ,\ 
    \alpha'_{ij}  \coloneqq -e^2 \frac{\partial^2 M_{ij}}{\partial \mu^2} 
    \label{eq:alpha_MQ},\\
    \alpha_{ijk}   &\coloneqq e   \frac{\partial M_{ijk}}{\partial \mu}     ,\ 
    \alpha'_{ijk}   \coloneqq e^2 \frac{\partial^2 M_{ijk}}{\partial \mu^2} ,\ 
    \alpha''_{ijk}  \coloneqq e^3 \frac{\partial^3 M_{ijk}}{\partial \mu^3}
    \label{eq:alpha_MO}.
\end{align}
As discussed in Refs.~\cite{gao:prb2018spinMQM,shitade:prb2018orbitalMQM,gao:prb2018orbitalMQM,shitade:prb2019spinMQM},
the magnetic multipoles characterize the charge polarization induced by the magnetic fields.
To proceed with the rigorous discussion, we restrict ourselves to 
insulators at $T=0$ here.
By using Eqs.~\eqref{eq:maxwell2}, \eqref{eq:maxwell3}, \eqref{eq:alpha_MQ}, and \eqref{eq:alpha_MO},
we can write the polarization charge induced by the magnetic field as
\begin{align}
    e \Delta N &= - \partial_{r_k} P^{\mathrm{tot}}_{k} , \quad
    P^{\mathrm{tot}}_{k} = \alpha_{ik} B_{i} - \alpha_{ijk} \partial_{r_j} B_{i} ,
    \label{eq:pola_charge}
\end{align}
where $\bm{P}^{\mathrm{tot}}$ is the total polarization
incorporating the contribution from the spatial variation of the magnetic field.
Equation~\eqref{eq:pola_charge} shows that $\alpha_{ij}$ corresponds to the magnetoelectric polarizability%
~\cite{essin:prl2009,essin:prb2010},
whereas $\alpha_{ijk}$ corresponds to the quadrupolar magnetoelectric coupling, 
which characterizes the charge polarization induced by a magnetic field gradient.
The quadrupolar magnetoelectric coupling may also characterize 
the magnetic field correction to EQs:
\begin{align}
    Q_{jk} = \alpha_{ijk} B_{i} ,
     \label{eq:Q_mag_correction}
\end{align}
because the spatial variation of the EQ contributes to the electric polarization
through the classical relation:
\begin{align}
    P^{\mathrm{tot}}_{k} = P_k - \partial_{r_j} Q_{jk} 
    \label{eq:PQrelation}.
\end{align}
Note that Eqs.~\eqref{eq:Q_mag_correction} and \eqref{eq:PQrelation} 
involve some ambiguity and therefore requires caution, 
because the electric polarization itself cannot be formulated within the thermodynamic framework~\cite{daido:prb2020}.
In addition, if one calculates $P^{\mathrm{tot}}_k$ up to the first-order spatial derivative with respect to the magnetic field, 
as in Eq.~\eqref{eq:pola_charge}, $\alpha_{ijk}$ in Eq.~\eqref{eq:pola_charge} 
is considered to include not only the magnetic-field correction to $Q_{jk}$ 
but also the contribution from the magnetic-field-gradient correction to $P_k$.
Equations~\eqref{eq:pola_charge} and \eqref{eq:Q_mag_correction} demonstrate that
the MQ relates to the magnetoelectric polarizability%
~\cite{gao:prb2018spinMQM,shitade:prb2018orbitalMQM,gao:prb2018orbitalMQM,shitade:prb2019spinMQM} and
the MO relates to the quadrupolar magnetoelectric coupling, or equivalently,
the correction to EQ by the magnetic fields.
Since the microscopic origin of the strain is EQ%
~\cite{hayami:prb2018classification,yatsushiro:prb2021classification,hayami:jpsj2024review},
we can consider that the MO is also related to the piezomagnetic effect~%
\cite{bhowal:prx2024,radaelli:prb2024}
as a representative response.

Moreover, the MQ and MO are related to the magnetization induced by electric fields~\cite{shitade:prb2018orbitalMQM}.
To see this explicitly, we express the total magnetization,
incorporating the contribution from the spatially varying MQ and MO, as
\begin{align}
    M^{\mathrm{tot}}_{i} 
    &= M_i
    -\partial_{r_j} M_{ij} + \partial_{r_j} \partial_{r_k} M_{ijk} ,
\end{align}
where $\bm{M}$ ($\bm{M}^{\mathrm{tot}}$) denotes the (total) magnetization.
Considering a nonuniform chemical potential $\mu(\bm{r})$, the total magnetization can be written as
\begin{align}
    M^{\mathrm{tot}}_{i} 
    &= M_i -\partial_{r_j} \mu \frac{\partial M_{ij}}{\partial \mu}
    + \partial_{r_k} \left[ \partial_{r_j} \mu \frac{\partial M_{ijk}}{\partial \mu} \right] \nonumber \\
    &= M_i -\partial_{r_j} \mu \frac{\partial M_{ij}}{\partial \mu} \nonumber \\
    &\hspace{1em}+ \partial_{r_j} \partial_{r_k} \mu \frac{\partial M_{ijk}}{\partial \mu}
    + \partial_{r_j} \mu \partial_{r_k} \mu \frac{\partial^2 M_{ijk}}{\partial \mu^2}
    \label{eq:magtot1} .
\end{align}
In the case of an insulator at $T=0$ in equilibrium, 
$\partial_{r_j} \mu / e$ is identified as an electric field $E_{j}$, hence we obtain
\begin{align}
    M^{\mathrm{tot}}_{i}
    &= M_i + \alpha_{ij} E_{j} + \alpha_{ijk} \partial_{r_j} E_{k}
    + \alpha'_{ijk} E_{j} E_{k}
    \label{eq:magtot2} ,
\end{align}
where $\alpha_{ij}$ characterizes the magnetization induced by an electric field 
(i.e., the magnetoelectric effect)~\cite{shitade:prb2019spinMQM,arai:jpsj2025}, 
$\alpha_{ijk}$ characterizes the magnetization induced by an electric field gradient~\cite{shitade:prb2025},
and $\alpha'_{ijk}$ characterizes the magnetization induced by a second-order electric field
(i.e., the second-order magnetoelectric effect)~\cite{xiao:prl2022hkspaceBCP, urru:annphys2022, oike:prb2024}.
On this basis, we can write the magnetization current for the orbital magnetic multipoles
$J^{\mathrm{tot}}_{a} = \epsilon_{a b i} \partial_{r_b} M^{\mathrm{tot}}_{i}$
induced by the electric fields in equilibrium and their spatial derivatives~\cite{shitade:prb2018orbitalMQM}:
\begin{align}
    J^{\mathrm{tot}}_{a} &= J^{\mathrm{D}}_{a} + J^{\mathrm{Q}}_{a} + J^{\mathrm{O}}_{a} , \\
    J^{\mathrm{D}}_{a}
    & = \epsilon_{a b i} \alpha_i E_b
     = \sigma_{ab} E_b
    \label{eq:Streda_MD},\\
    J^{\mathrm{Q}}_{a}
    &= \epsilon_{a b i} \left( \alpha_{ij} \partial_{r_b} E_j + \alpha'_{ij} E_b E_j \right) \nonumber \\
    &= \sigma_{abj} \partial_{r_b} E_j + \sigma'_{abj} E_b E_j
    \label{eq:Streda_MQ},\\
    J^{\mathrm{O}}_{a}
    &= \epsilon_{a b i} \Big\{ \alpha_{ijk} \partial_{r_b} \partial_{r_j} E_k + \alpha''_{ijk} E_b E_j E_k \nonumber \\
    &\hspace{2em} + \alpha'_{ijk} ( \partial_{r_b} E_j E_k + \partial_{r_k} E_b E_j + \partial_{r_j} E_k E_b ) \Big\} \nonumber \\
    &= \sigma_{abjk} \partial_{r_b} \partial_{r_j} E_k + \sigma''_{abjk} E_b E_j E_k \nonumber \\
    &\hspace{2em} + \sigma'_{abjk} ( \partial_{r_b} E_j E_k + \partial_{r_k} E_b E_j + \partial_{r_j} E_k E_b )
    \label{eq:Streda_MO},
\end{align}
where $\epsilon_{abi}$ is the totally antisymmetric tensor and
$J^{\mathrm{D}}_{\mu}$, $J^{\mathrm{Q}}_{\mu}$, and $J^{\mathrm{O}}_{\mu}$ are the magnetization current
arising from the orbital MD (magnetization), the orbital MQ, and the orbital MO, respectively.
The charge conductivity tensors are defined as follows:
\begin{align}
    \sigma_{ab}     &\coloneqq \epsilon_{a b i} \alpha_i       \label{eq:sigma_ab},\\
    \sigma_{abj}    &\coloneqq \epsilon_{a b i} \alpha_{ij}    \label{eq:sigma_abj},\ 
    \sigma'_{abj}    \coloneqq \epsilon_{a b i} \alpha'_{ij}   ,\\
    \sigma_{abjk}   &\coloneqq \epsilon_{a b i} \alpha_{ijk}    \label{eq:sigma_abjk},\ 
    \sigma'_{abjk}   \coloneqq \epsilon_{a b i} \alpha'_{ijk}   ,\ 
    \sigma''_{abjk}  \coloneqq \epsilon_{a b i} \alpha''_{ijk}  .
\end{align}
$\sigma_{ab}$ corresponds to the anomalous Hall conductivity (AHC), 
and Eqs.~\eqref{eq:sigma_ab} gives
the St\v{r}eda formula~\cite{streda:jphysc1982,widom:physletta1982,ceresoli:prb2006,xiao:rmp2010}:
\begin{align}
    \sigma_{ab} 
    = e \epsilon_{abi} \frac{\partial M_i}{\partial \mu} 
    = e \epsilon_{abi} \frac{\partial N}{\partial B_i} ,
\end{align}
where we used the Maxwell relation~\eqref{eq:maxwell1}.
Similarly, the first equations in Eqs.~\eqref{eq:sigma_abj} and \eqref{eq:sigma_abjk} give higher-rank analogues of the St\v{r}eda formula for the orbital MQ and MO, respectively:
\begin{align}
    \sigma_{abj} 
    &= -e \epsilon_{abi} \frac{\partial M_{ij}}{\partial \mu} 
     = -e \epsilon_{abi} \frac{\partial N}{\partial [\partial_{r_j} B_i]} , \\
    \sigma_{abjk} 
    &= e \epsilon_{abi} \frac{\partial M_{ijk}}{\partial \mu} 
    = e \epsilon_{abi} \frac{\partial N}{\partial [\partial_{r_j} \partial_{r_k} B_i]} .
\end{align}
We refer to $\sigma_{abj}$ as quadrupolar AHC~\cite{essin:prb2010, malashevich:prb2010},
whereas $\sigma'_{abj}$ as second-order AHC~\cite{gao:prl2014BCP,das:prb2023NLAHE,kaplan:prl2024NLAHE}.
In addition, we refer to $\sigma_{abjk}$, $\sigma'_{abjk}$, and $\sigma''_{abjk}$
as octupolar AHC~\cite{kozii:prl2021}, second-order quadrupolar AHC, and third-order AHC~\cite{xiang:prb2023}, respectively.
Here, we only consider the static responses, with the electric field satisfying $\partial_{r_i} E_j = \partial_{r_j} E_i$.
We consider that these higher-rank and higher-order AHC tensors 
are antisymmetric under the exchange of the first and the other indices,
e.g., $\sigma_{abj}=-\sigma_{baj}$ and $\sigma_{abjk}=-\sigma_{bajk}$,
and symmetric under the exchange among the second and above indices,
e.g., $\sigma_{abj}=\sigma_{ajb}$ and $\sigma_{abjk}=\sigma_{ajbk}$.
Note that these conductivity tensors arise from intrinsic origins and are free from dissipation.

The above discussion demonstrates that magnetic multipoles are directly related to physical responses.
However, as initially assumed, this argument is strictly valid only for insulators at $T=0$.
Consequently, regarding 
the MQ and MO, response tensors expressed in terms of 
higher-order derivatives with respect to the chemical potential, 
i.e., tensors containing $\alpha'_{ij}$, $\alpha'_{ijk}$, and $\alpha''_{ijk}$, 
have no direct relation to the multipoles within our thermodynamic framework.
This is because, in insulators at $T=0$, 
the following substitutions hold in the expressions for the MQ and MO: 
$\sum_n f_n \rightarrow \sum_{n}^{\mathrm{occ.}}$ and
$-\sum_n \int_{\epsilon_n-\mu}^{\infty} dz f(z) \rightarrow \sum_{n}^{\mathrm{occ.}} (\epsilon_n-\mu)$,
which makes MQ and MO linear functions of the chemical potential%
~\cite{gao:prb2018spinMQM, shitade:prb2018orbitalMQM, gao:prb2018orbitalMQM, shitade:prb2019spinMQM}.

Under multipole orderings, the rank and spacetime symmetry determine which response tensors may be finite%
~\cite{hayami:prb2018classification, watanabe:prb2018classification, yatsushiro:prb2021classification, hayami:jpsj2024review,
spaldin:jpcm2008, watanabe:prb2017, radaelli:prb2024, mcclarty:prl2024, schiff:prr2025}.
In Table~\ref{tab:response}, we summarize the correspondence 
between magnetic multipoles in crystalline solids and the response tensors 
that can become finite upon the ordering from the symmetry viewpoint.
Based on thermodynamic considerations in insulators at $T=0$, we further classify these tensors into three types:
\begin{enumerate}
    \item[(I)]   Those that are obtained from the first-order derivative of the magnetic multipole with respect to $\mu$
    (i.e., tensors containing $\alpha_i$, $\alpha_{ij}$, or $\alpha_{ijk}$).
    \item[(II)]  Those that are obtained from the higher-order derivative of the magnetic multipole with respect to $\mu$
    (i.e., tensors containing $\alpha'_{ij}$, $\alpha'_{ijk}$, or $\alpha''_{ijk}$).
    \item[(III)] Those that do not appear in the thermodynamic arguments.
\end{enumerate}
Type~I (II) responses are always linear (nonlinear) responses.
Type~II responses in the MD ordering do not appear in our arguments.
Dissipative responses are always classified into Type~III
because such effects are not incorporated within our thermodynamic framework.
As representative examples, we list the spin current generation
$J_i \sigma_j = \sigma^{\mathrm{S}}_{ijk} J_k$,
where $J_i \sigma_j$ ($J_k$) represents the spin (charge) current, for the MD and MO orderings,
and the current-induced distortion
$\zeta_{ij} = d_{ijk} J_k$, where $\zeta_{ij} = \partial u_i / \partial r_j$ and $u_i$ is the displacement,
for the MQ orderings in Table~\ref{tab:response}.
Note that these responses are accompanied by dissipation.
We also list the (intrinsic) AHE for the MO orderings as a Type~III response.
We focused on the static responses here
since we relate the multipoles to the responses in terms of the thermodynamic arguments. 
Therefore, dynamical responses such as optical responses are not included in Table~\ref{tab:response}. 
It should also be noted that the Type~III responses listed in the table represent only a subset and are not exhaustive.

The present analysis through the thermodynamic relations does not imply that the Type~II and Type~III responses 
are completely unrelated to the multipoles. 
Effects that are not captured within our thermodynamic framework, 
such as finite temperature, nonequilibrium state, dissipation, finite relaxation time, time-dependent fields, 
or contributions not originating from spatially varying MQ and MO, may play a certain role.
We believe that the present classification will also be useful in such general situations.
For example, in a metallic system, the replacement $\partial_{r_j} \mu / e \rightarrow E_j$ 
in Eq.~\eqref{eq:magtot2} is not strictly accurate, but it retains a certain physical meaning 
if we consider the response in a nonequilibrium state.
Furthermore, it is also an important direction to consider
thermoelectric/magnetic responses by introducing spatially inhomogeneous temperature $T(\bm{r})$ 
instead of the inhomogeneous chemical potential $\mu(\bm{r})$%
~\cite{xiao:prl2006thermoele, xiao:rmp2010, gao:prb2018orbitalMQM, shitade:prb2019spinMQM}.
Actually, it has been demonstrated microscopically that the orbital MD and MQ give rise to 
linear and nonlinear anomalous thermoelectric currents, respectively~\cite{xiao:prl2006thermoele,gao:prb2018orbitalMQM}.
In addition, it should also be noted that the magnetization current itself 
is not a directly measurable quantity in conventional transport experiments, 
which probe nonequilibrium states~\cite{cooper:prb1997,xiao:prl2006thermoele,gao:prb2018orbitalMQM}.
While it has been established that $\sigma_{ab}$ can be measured experimentally, 
the observability of the higher-rank conductivities, $\sigma_{abj}$ and $\sigma_{abjk}$, 
requires further confirmation through microscopic calculations.
A detailed investigation of how these factors connect the responses and the multipoles remains an important task.

\subsection*{Application to collinear magnets}
In this subsection, we perform model calculations for the obtained MO.
We consider the two-sublattice model of a time-reversal-symmetry breaking (TRSB) AFM in the presence of the spatial inversion symmetry~\cite{roig:prb2024}:
\begin{align}
    \hat{H}^{\mathrm{Para}}_{\bm{k}}
    &= \varepsilon_{0,\bm{k}} + t_{x,\bm{k}} \hat{\tau}_{x} + t_{z,\bm{k}} \hat{\tau}_{z}
    + \hat{\tau}_{y} \bm{\lambda}_{\bm{k}} \cdot \hat{\bm{\sigma}}
    \label{eq:Hk_for_Para} , \\
    \hat{H}^{\mathrm{AFM}}_{\bm{k}} &=
    \hat{H}^{\mathrm{Para}}_{\bm{k}} + \hat{\tau}_{z} \bm{J} \cdot \hat{\bm{\sigma}} ,
    \label{eq:Hk_for_AFM}
\end{align}
where $\hat{H}^{\mathrm{Para}}_{\bm{k}}$ and $\hat{H}^{\mathrm{AFM}}_{\bm{k}}$ represent the Hamiltonian
in the paramagnetic and antiferromagnetic states, respectively.
$\hat{\tau}_{i}$ and $\hat{\sigma}_{i}$ are the Pauli matrices for the sublattice and spin space,
$t_{x,\bm{k}}$ and $t_{z,\bm{k}}$ are inter- and intrasublattice hopping,
$\bm{\lambda}_{\bm{k}} = \lambda \tilde{\bm{\lambda}}_{\bm{k}}$ represents spin-orbit coupling (SOC), and
$\bm{J}$ is the magnetic moment localized at each sublattice. 
We here focus on a collinear spin configuration, although the expression of the MO can be applied to a noncollinear one.
We use model parameters that reproduce the nonmagnetic band structures
for MnF$_2$ listed in Ref.~\cite{roig:prb2024},
where the point group symmetry in Eq.~(\ref{eq:Hk_for_Para})
without SOC is $D_{4h}$;
we adopt the lattice constant as the length unit.
The details of the model
are also shown in Methods section.

\begin{figure}[tbp]
    \includegraphics[width=8.5cm]{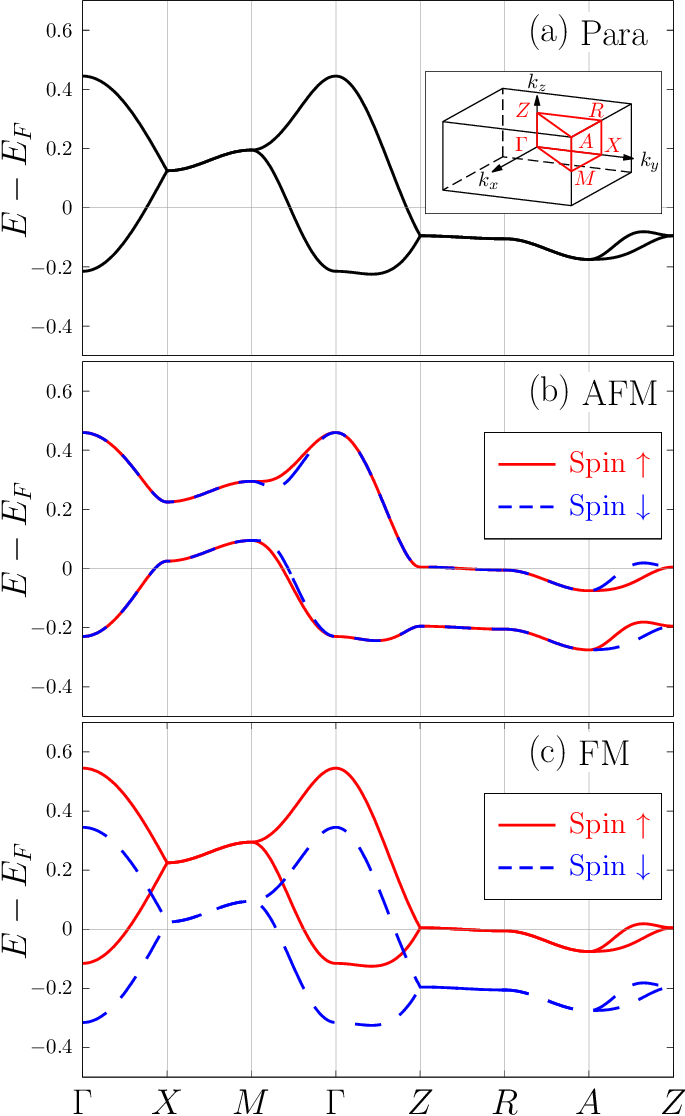}
    \caption{
        \textbf{Electronic band dispersions of the collinear magnetic models.}
        Band structures in the paramagnetic (a), the antiferromagnetic (b), and
        the ferromagnetic states (c) without SOC calculated from the Hamiltonian
        in Eqs.~\eqref{eq:Hk_for_Para}, \eqref{eq:Hk_for_AFM}, and \eqref{eq:Hk_for_FM}, respectively.
        The parameters are chosen to reproduce
        the paramagnetic band structures of MnF$_2$ obtained
        in Ref.~\cite{roig:prb2024} (see also Methods section
        for details)
        and $J=0.1$ in (b) and (c).
    }
    \label{fig:band}
\end{figure}

In Fig.~\ref{fig:band}(a), we show the band structures of
the paramagnetic state ($\hat{H}^{\mathrm{Para}}_{\bm{k}}$)
without SOC, i.e., $\lambda=0$~\cite{setyawan:compmatsci2010,roig:prb2024}.
There are two spin-degenerate bands due to time-reversal and spatial inversion symmetry.
The corresponding band structures of the antiferromagnetic state ($\hat{H}^{\mathrm{AFM}}_{\bm{k}}$)
at $J=0.1$ are shown in Fig~\ref{fig:band}(b);
the up- and down-spin bands are shown
by the red solid lines and blue dashed lines, respectively.
We obtain the spin splitting along the $M-\Gamma$ and $A-Z$ lines
originating from the intrasublattice hopping and the magnetic moment.
$\hat{H}^{\mathrm{AFM}}_{\bm{k}}$ describes the electronic structures of the so-called
$d_{xy}$-wave altermagnet~\cite{roig:prb2024}.

Together with the above model, we also consider the model of an FM:
\begin{align}
    \hat{H}^{\mathrm{FM}}_{\bm{k}} &=
    \hat{H}^{\mathrm{Para}}_{\bm{k}} + \bm{J} \cdot \hat{\bm{\sigma}} .
    \label{eq:Hk_for_FM}
\end{align}
Figure~\ref{fig:band}(c) shows the band structures of the FM with $J=0.1$ and $\lambda=0$.
The model shows the isotropic ($s$-wave) spin splitting, and hence has net magnetization.

\begin{figure}[tbp]
    \includegraphics[width=8.5cm]{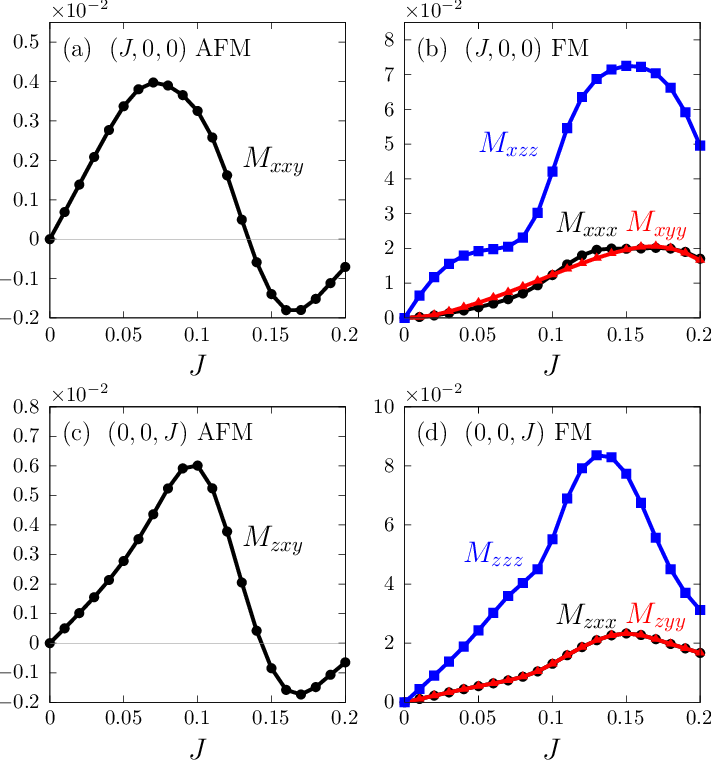}
    \caption{
        \textbf{Finite components of $M_{ijk}$ as a function of $J$.}
        [(a) and (b)] ([(c) and (d)])
        $J$ dependence of the
        nonzero components of $M_{ijk}$ for $\bm{J}=(J,0,0)$ ($\bm{J}=(0,0,J)$) in the AFM and FM, respectively.
        We choose $\lambda=0.1$ and $T=0.01$.
    }
    \label{fig:Mijk_jdep}
\end{figure}

Figure~\ref{fig:Mijk_jdep} shows the $J$ dependence of the nonzero components of the MO
for the collinear magnets with SOC.
In (a) and (b), we show the results for $\bm{J}=(J,0,0)$
in the AFM and FM, respectively.
The results for $\bm{J}=(0,0,J)$ in the AFM and FM are shown in (c) and (d).
We assumed that the SOC has the same orientation as the magnetic moment
($\bm{\lambda}_{\bm{k}} \parallel \bm{J}$) and $\lambda = 0.1$.
In this situation, the orientation of the spin is not canted from the direction of $\bm{J}$.
The temperature is set to $0.01$ in the numerical simulations.
In the collinear magnets,
the first index of nonzero $M_{ijk}$ reflects the direction of the magnetic moment in the Hamiltonian
and the second and third indices reflect
the symmetry of the spin splitting such as $d_{xy}$- and $s$-wave.
When the amplitude of $\bm{J}$ is small, $M_{ijk}$ varies monotonically.
In the vicinity of $J \sim 0.1$, on the other hand, $M_{ijk}$ shows non-monotonic behavior.
It is not easy to clarify the reasons
because the $J$ dependence is strongly influenced by the details of the band structures.
We also show the $J$
dependence of the MO for $\lambda=0$
in Supplemental Material~\cite{supple}, and 
find qualitatively the same results as those for $\lambda=0.1$.
We find that the sign change in these cases is associated with 
a deformation of the Fermi surfaces.
A detailed discussion is provided in Supplemental Material~\cite{supple}.
We also calculated the MO by using the parameters
for the one-orbital model of RuO$_2$
obtained in Ref.~\cite{roig:prb2024} and present the results in Supplemental Material~\cite{supple}.
We chose model parameters relevant to RuO$_2$
to examine the properties of the MO under different parameter settings.
Whether these parameters accurately represent the electronic structure of real materials
is beyond the scope of the present work and will not be discussed here,
although the actual magnetic structure of RuO$_2$ is still under debate~\cite{hiraishi:prl2024}.

\begin{figure}[tbp]
    \includegraphics[width=8.5cm]{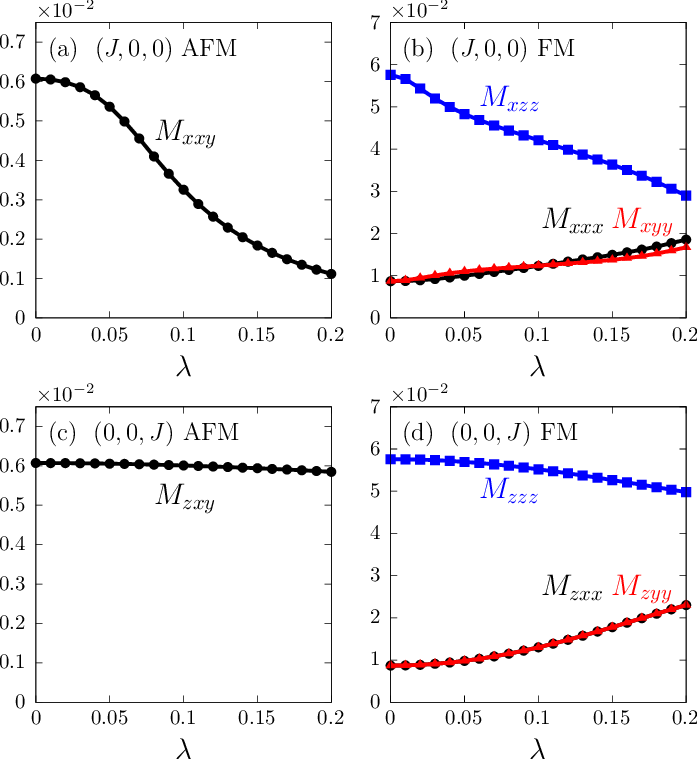}
    \caption{
        \textbf{Finite components of $M_{ijk}$ as a function of $\lambda$.}
        [(a) and (b)] ([(c) and (d)])
        $\lambda$ dependence of the
        nonzero components of $M_{ijk}$ when $\bm{J}=(J,0,0)$ ($\bm{J}=(0,0,J)$) in the AFM and FM, respectively.
        We choose $J=0.1$ and $T=0.01$.
    }
    \label{fig:Mijk_lamdep}
\end{figure}

Figure~\ref{fig:Mijk_lamdep} shows the $\lambda$ dependence of the nonzero components of the MO
with $\bm{\lambda}_{\bm{k}} \parallel \bm{J}$, $J=0.1$, and $T=0.01$.
In (a) and (b), we show the results for $\bm{J}=(J,0,0)$
in the AFM and FM, respectively.
The results for $\bm{J}=(0,0,J)$ in the AFM and FM are shown in (c) and (d).
Compared with the dependence on $J$ in Fig.~\ref{fig:Mijk_jdep}, there is little dependence on $\lambda$.
This result is a consequence of the SOC, which is parallel to the magnetic moment,
having little effect on the magnetic structures.
The MO captures the nonrelativistic properties of the model well.
The results for the one-orbital model of RuO$_2$~\cite{roig:prb2024}
are also shown in Supplemental Material~\cite{supple}
and similar results are obtained. 

\begin{figure*}[tbp]
    \includegraphics[width=17.5cm]{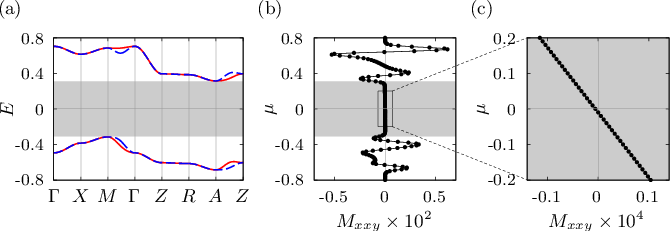}
    \caption{
        \textbf{Chemical potential dependence of the MO.}
        (a) Antiferromagnetic band structure of the collinear model with an insulating gap.
        Unlike Fig.~\ref{fig:band}, the offset due to the Fermi level is not taken into account.
        (b) $M_{xxy}$ is plotted as a function of the chemical potential.
        (c) Enlarged view of the boxed area in (b).
        In (a)-(c), the parameters are chosen to be as follows: 
        $\bm{J}=(J,0,0)$, $\bm{\lambda}_{\bm{k}}=(\lambda_{x,\bm{k}},0,0)$, $J=0.5$, and $\lambda=0.1$.
        We choose $T=0.01$ in (b) and (c).
        The shaded areas correspond to the energy gap.
    }
    \label{fig:Mijk_mudep}
\end{figure*}

Next, we investigate the thermodynamic relation of the MO.
We calculate the $\mu$ dependence of the nonzero components of the MO 
in the collinear antiferromagnetic state with $\bm{J}=(J,0,0)$ and $\bm{\lambda}_{\bm{k}} \parallel \bm{J}$.
To obtain an insulating region, we set $J=0.5$. 
$\lambda$ is set to 0.1.
In addition, we choose $T=0.01$, much smaller than the insulating gap ($E_g \sim 0.6$), 
to avoid technical difficulties in numerical calculations.
The electronic band structure of the insulating AFM is shown in Fig.~\ref{fig:Mijk_mudep}(a) and 
the $\mu$ dependence of the finite component of the MO is shown in (b) and (c).
We confirm the linear dependence of the MO on $\mu$ within the insulating gap,
supporting the discussion in the previous subsection.
The finite slope is indicative of the corresponding responses shown in Table.~\ref{tab:response}.
We also note that the slope may vary even if the parameters of the Hamiltonian are changed 
while keeping the insulating gap open, implying that there is no topological constraint in this situation.

\section*{Discussion}
\begin{figure}[tbp]
    \includegraphics[width=8.5cm]{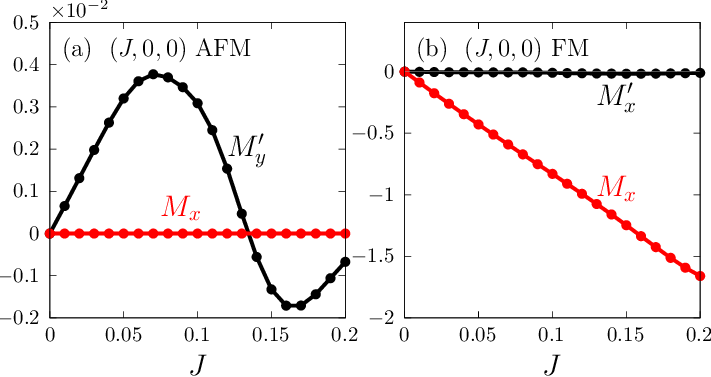}
    \caption{
        \textbf{Magnetic dipoles as a function of $J$.}
        $J$ dependence of the finite components of the AMD ($M'_{i}$) and the spin magnetization ($M_{i}$)
        in the AFM (FM) is shown in (a) ((b)), where the vertical axis is expressed in units of $\mu_{\mathrm{B}}$.
        Although the spin magnetization remains zero in the AFM, it is shown for reference.
        The parameters are the same as those in Fig.~\ref{fig:Mijk_jdep}.
    }
    \label{fig:AMD}
\end{figure}

As discussed in Results section,
$M_{ijk}$ is a reducible tensor
and we focus on the contribution of AMD here.
In Fig.~\ref{fig:AMD}, we compare the $J$ dependence of the AMD $M'_{i}$
with that of the spin magnetization $M_{i}$ [Eq.~\eqref{eq:MD}] in
both AFM and FM systems.
As in the previous section, we choose
$\bm{J}=(J,0,0)$, $\bm{\lambda}_{\bm{k}} \parallel \bm{J}$, $\lambda=0.1$, and $T=0.01$.
In the AFM, $M'_y=3M_{xxy}/\sqrt{10}$ becomes finite when $J \neq 0$, while $M_{i}=0$ ($i=x,y,z$).
In the FM, on the other hand, $M_x$ increases almost linearly with increasing $J$,
while $M'_x=(2M_{xxx}-M_{xyy}-M_{xzz})/\sqrt{10}$ appears nearly unchanged and negligibly small compared to $M_x$.
Therefore, the MD that characterizes the magnetic structure of the AFM (FM)
with $\bm{J}=(J,0,0)$ is identified as $M'_y$ ($M_x$).
It is also worth mentioning that AMD corresponds to a $\bm{T}$ vector,
which is an experimentally observable quantity by the measurement of
the x-ray magneto-circular dichroism (XMCD)~\cite{carra:prl1993,stohr:prl1995,stohr:jesrp1995,crocombette:jpcm1996,yamasaki:jpsj2020}.
When the magnetic moment is oriented along the $z$-axis with $\bm{J}=(0,0,J)$, on the other hand,
the only nonzero component of $M_{ijk}$ in the AFM is $M_{zxy}$.
This component contributes to MO $M_{[xyz]}=M_{zxy}/3$ and MTQ $T_{v}=M_{zxy}$~\cite{supple}
and does not contribute to AMD.
We confirm that these properties are valid for other parameter settings
(see Supplemental Material~\cite{supple} for detail).

This result is consistent with the following symmetry analysis.
The TRSB terms
in Eqs.~\eqref{eq:Hk_for_AFM} and \eqref{eq:Hk_for_FM} often break mirror symmetries
and allow the axial vectors to exist~\cite{roig:prb2024}.
In Table~\ref{tab:axial}, we summarize the TRSB term,
the symmetry-allowed axial vector $\bm{h}=(h_x, h_y, h_z)$ under the point group $D_{4h}$, and
dominant MD in each model including the FM.
Focusing on the AFM, $\hat{\tau}_z \hat{\sigma}_x$ and $\hat{\tau}_z \hat{\sigma}_y$ allow
nonzero $h_y$ and $h_x$, respectively.
$\hat{\tau}_z \hat{\sigma}_z$, on the other hand, prohibits a finite $\bm{h}$.
Therefore, depending on the orientation of the magnetic moment in the Hamiltonian,
the existence of an MD is allowed
even in the AFM without the spin magnetization.
This MD is nothing but the AMD~\cite{kusunose:jpsj2020,hayami:prb2021AMD}.

Moreover, $\hat{\tau}_z \hat{\sigma}_x$ and $\hat{\tau}_z \hat{\sigma}_y$
make the Hall conductivity tensor finite~\cite{roig:prb2024}.
Since the microscopic origin of the Hall conductivity tensor
lies in MDs~\cite{hayami:prb2018classification,yatsushiro:prb2021classification,hayami:jpsj2024review},
the fundamental origin of the AHE in the AFMs can be considered as the AMD~\cite{hayami:prb2021AMD,ohgata:prb2025}.
Notably, this AMD-driven AHE does not necessarily require 
a finite spin magnetization arising from SOC-induced spin canting~\cite{smejkal:sciadv2020,naka:prb2020,hayami:prb2021AMD}.
In real materials with complex electronic structures, however, 
both the AMD and the spin magnetization arising from the spin canting are expected to be present. 
Therefore, it may not be easy to distinguish the contributions to the AHE from the AMD and from the spin magnetization. 
However, we believe that XMCD provides useful information to make this distinction.
We also summarize the nonzero component of the AHC $\sigma_{ij}$ 
in the model [Eqs.~\eqref{eq:Hk_for_AFM} and \eqref{eq:Hk_for_FM}]
in the fourth row of Table~\ref{tab:axial}~\cite{roig:prb2024}.
Existence of the AHE clearly illustrates the difference between the antiferromagnetic state with 
$\hat{\tau}_z \hat{\sigma}_x$ ($\hat{\tau}_z \hat{\sigma}_y$) and that with $\hat{\tau}_z \hat{\sigma}_z$.
The latter state can, for example, be illustrated by the spin current generation.
It should be noted that both the AHE and the spin current generation correspond to the Type~III response 
in the MO ($M_{ijk}$) orderings, as discussed in Results section. 
Accordingly, within the scope of the present discussion, we cannot analyze it beyond symmetry considerations.

To discuss the connection between the Type~III response and the MO beyond symmetry arguments,
we comment here on the AHE in the collinear antiferromagnetic model from a microscopic viewpoint.
The intrinsic AHC is obtained by integrating the Berry curvature $\Omega_n^{ij}$
over the occupied electronic states~\cite{karplus:pr1954AHE, nagaosa:rmp2010AHE}:
\begin{align}
    \sigma_{ij} &= -\frac{e^2}{\hbar} \sum_n \int \frac{d^d k}{(2\pi)^d}\, \Omega_n^{ij}\, f_n
    \label{eq:AHC} ,
\end{align}
where $\hbar$ is explicitly written following the standard convention.
In the collinear antiferromagnetic model [Eq.~\eqref{eq:Hk_for_AFM}] with $\bm{J}=(J,0,0)$ or $(0,J,0)$,
the finite SOC gives rise to a nonzero Berry curvature~\cite{roig:prb2024},
whereas a finite MO does not require SOC, as shown in Results section.
Moreover, the Berry curvature corresponds to the imaginary part of the QGT,
$\Omega_n^{ij} = -2\mathrm{Im}\, T_n^{ij}$,
while the MO involves the real part of the geometric quantities such as the QM,
$g_n^{ij} = \mathrm{Re}\, T_n^{ij}$.
These considerations indicate that the relationship between the AHE and the MO (AMD) is likely indirect. 
For further reference, we show the momentum-space distributions of the AHC
and the MO in the Supplemental Material~\cite{supple}. 
Another Type~III response in the MO ordering is the spin current generation, which does not require SOC~\cite{naka:natcomm2019}. 
In this respect, the relationship between the MO and the spin current generation 
is expected to differ from that of the AHE. 
However, since the spin current generation is a dissipative response, 
it remains unclear whether it has a direct connection to the MO, which is an equilibrium quantity. 
A comprehensive understanding of the relationship between Type~III responses and the MO deserves further study.

Our gauge-independent MO is more informative and broadly applicable than the N\'{e}el vector, although 
both share the same symmetry properties.
The MO can be directly applied to a wide range of magnetic systems, including noncollinear AFMs,
as long as the periodic Hamiltonian is defined, whereas the N\'{e}el vector is not always unambiguously defined.
Moreover, unlike the N\'{e}el vector, the MO reflects the underlying electronic structure 
and is directly connected to physical responses via thermodynamic relations beyond symmetry considerations.
These features make it a useful tool for characterizing unconventional TRSB magnets.
However, it is important to note that our MO should not be regarded as identical 
to the order parameter used in Landau free energy expansions 
and related contexts~\cite{mcclarty:prl2024, schiff:prr2025}.
The Landau order parameter is defined purely by symmetry.
It is always finite once the symmetry is broken and is valid only near the transition temperature 
where the order parameter remains small.
In contrast, our MO incorporating electronic structure information
can change its sign depending on microscopic details, 
and remains valid across the entire temperature range.
In this paper, we have not fully explored the effects of temperature.
We consider that the primary effect of temperature is to influence the MO 
through changes in the magnitude of the mean field~\cite{daido:prb2020}.
Although the effects of temperature may differ between the Fermi-surface terms 
and the other terms in $M_{ijk}$, we expect the aforementioned effect to be the most significant.
For reference, we present the results of the $J$ dependence calculated at different temperatures 
for the collinear AFM in Supplemental Material~\cite{supple}.
A detailed investigation of the temperature dependence 
will be addressed elsewhere, 
particularly in calculations based on realistic and complex electronic structures.

It is not straightforward to obtain a clear understanding of the relation 
to the locally defined MO~\cite{hayami:prb2021AMD,bhowal:prx2024}.
Here, we note that the locally defined MO seems to lack the itinerant contribution 
represented by the terms proportional to the grand potential density 
(integration of the Fermi distribution function) in Eq.~\eqref{eq:Mijk}%
~\cite{xiao:prl2005, xiao:prl2006thermoele, thonhauser:prl2005, ceresoli:prb2006}.
We also remark that the itinerant terms, representing contributions from the edge, remain finite even in insulators.
We believe that our MO offers complementary insights to those presented 
in Refs.~\cite{hayami:prb2021AMD,bhowal:prx2024}
and serves as a better bulk order parameter than the local one in the thermodynamic limit.
Moreover, prior studies could only infer possible responses from symmetry considerations, 
our approach establishes a direct thermodynamic relation 
between the MO and measurable response functions, thereby providing deeper physical insight.
On the contrary, our MO may not be able to be applied to finite small samples but the local MO may be.
Combining these perspectives paves the way for a more comprehensive understanding of unconventional magnets.

In this paper, we focus on collinear TRSB AFMs with $d$-wave spin splitting, 
but our MO can be applied to a wide range of magnets, including noncollinear AFMs.
Therefore, application of the MO to such magnets is an important direction for future studies.
As a representative and simple example, we calculate the MO of the MnF$_2$ model 
in the case where $\bm{\lambda}_{\bm{k}}$ and $\bm{J}$ are not parallel.
In this situation, spin canting is induced by SOC, resulting in a noncollinear magnetic structure.
Consequently, additional components in $M_{ijk}$ become finite. 
We provide the results and discussions in Supplemental Material~\cite{supple}.
Furthermore, our MO may be able to describe the MO moment decomposed into sublattice contributions.
This decomposition may be useful for characterizing magnetic structures 
in which local MO moments have opposite signs on different sublattices.
Extending our formula to such antiferrooctupole order is also an important direction to explore.
In addition, the characterization of unconventional magnets 
with higher-order spin splitting, such as $g$-wave altermagnets, is an important issue.
In such orderings, the MO may have finite components, 
but higher-rank magnetic multipoles, such as magnetic triakontadipoles, 
may also be important for characterizing the magnetic structure~\cite{verbeek:prr2024}.
We also note that our MO can characterize TRSB magnets 
without inversion symmetry when the MO moment is allowed.
However, magnets in which the MO moment is prohibited, 
such as $\mathcal{PT}$-symmetric magnets, 
cannot be characterized by the MO.

In summary, we have formulated the gauge-invariant spin MO in crystalline solids
based on quantum mechanics and thermodynamics.
The direct relationships between the MO and several response tensors have been established in insulators at $T=0$.
We have classified the allowed physical responses into three types beyond symmetry considerations.
The properties of the obtained MO have been discussed by performing model calculations.
We conclude that the MO effectively characterizes the nonrelativistic properties of TRSB AFMs.
In particular, we confirm that the AMD, derived from the tensor decomposition of the MO and
experimentally observable in XMCD measurements, well describes the magnetic properties of
AFMs exhibiting AHE in the $d$-wave altermagnet model.
However, in this work, direct connections between the AMD and the AHE 
are not demonstrated microscopically beyond symmetry arguments.
Further investigation into the relationships between the AMD identified in this study and the AHE, 
as well as ferromagnetism-related physical responses in AFMs 
such as the magneto-optical Kerr and Nernst effects%
~\cite{hayami:prb2018classification, yatsushiro:prb2021classification, hayami:jpsj2024review,
smejkal:sciadv2020, smejkal:prx2022spingroup1, smejkal:prx2022spingroup2, solovyev:prb1997, shao:prapp2021, samanta:jappphys2020,
sivadas:prl2016, naka:prb2020, naka:prb2021AHEperovskite, attias:prb2024, chen:prb2022, smejkal:natrevmat2022, radaelli:prb2024,
mcclarty:prl2024, schiff:prr2025} remains an open issue.

\section*{Methods}
\subsection*{Expression of MO near the band touching point}
Equation~\eqref{eq:Mijk} is valid when all bands are isolated with each other.
In the presence of band touchings, careful treatment is required, as
the geometric terms in Eqs.~\eqref{eq:g}-\eqref{eq:Y} may diverge.
In this subsection, we show the expression of the MO in the presence of the degeneracy points~\cite{daido:prb2020}.
To derive the expression, we begin with an equivalent form of Eq.~\eqref{eq:Mijk}:
\begin{widetext}
\begin{align}
    M_{ijk} = g \mu_{\mathrm{B}} \sum_{n} \int \frac{d^d k}{(2\pi)^d}
    \Bigg( &\frac{1}{12} s^i_n v^j_n v^k_n f''_n
    + \sum_{m}^{\neq n}
    \left[ \frac{1}{4} (s^i_n+s^i_m) \mathrm{Re} \left[ v^j_{nm} v^k_{mn} \right] \mathcal{P}_{nm}
    + \frac{1}{4} \left\{ \mathrm{Re} \left[ s^i_{nm} v^j_{mn} \right] v^k_n + (j \leftrightarrow k) \right\} \mathcal{Q}_{nm}
    \right] \nonumber \\
    &+ \sum_{m}^{\neq n} \sum_{\ell}^{\neq n,m}
    \frac{-1}{2} \mathrm{Re} \left[ s^i_{nm} v^j_{m\ell} v^k_{\ell n} \right] \mathcal{R}_{nm\ell}
    \Bigg) ,
    \label{eq:Mijk2}
\end{align}
\end{widetext}
where
\begin{align}
    \mathcal{P}_{nm} &= \frac{1}{\epsilon_{nm}^2} \left( f_n + f_m - \frac{2\int_{nm}}{\epsilon_{nm}} \right) , \\
    \mathcal{Q}_{nm} &= \frac{1}{\epsilon_{nm}}
    \left\{ f'_n + 2\left( \frac{-f_n}{\epsilon_{nm}} + \frac{\int_{nm}}{\epsilon_{nm}^2} \right) \right\} , \\
    \mathcal{R}_{nm\ell} &= \frac{1}{\epsilon_{m\ell} \epsilon_{\ell n}}
    \left( f_{\ell} + \frac{\int_{nm}}{\epsilon_{nm}}
    - \frac{\int_{m\ell}}{\epsilon_{m\ell}} - \frac{\int_{\ell n}}{\epsilon_{\ell n}} \right) .
\end{align}
Here, we introduce the abbreviation $\int_{nm} = \int_{\epsilon_m - \mu}^{\epsilon_n - \mu} dz f(z)$.
When band $n$ is degenerate with band $m$ at $\bm{k}=\bm{k}_0$ ($\epsilon_{n\bm{k}_0} = \epsilon_{m\bm{k}_0}$),
$\mathcal{P}_{nm}$ and $\mathcal{Q}_{nm}$
have the asymptotic form which is valid near $\bm{k}=\bm{k}_0$:
\begin{align}
    \mathcal{P}_{nm} &= \frac{1}{6} f''_m + \frac{1}{12} f'''_m \epsilon_{nm} + \frac{1}{40} f''''_m \epsilon_{nm}^2 + O(\epsilon_{nm}^3)
    \label{eq:P_expand} , \\
    \mathcal{Q}_{nm} &= \frac{1}{3} f''_m + \frac{1}{4} f'''_m \epsilon_{nm} + \frac{1}{10} f''''_m \epsilon_{nm}^2 + O(\epsilon_{nm}^3)
    \label{eq:Q_expand} .
\end{align}
For $\mathcal{R}_{nm\ell}$, it is necessary to classify cases according to the type of degeneracy:
\begin{itemize}
    \item[(i)]   $\epsilon_{n\bm{k}_0} = \epsilon_{m\bm{k}_0}$ and $\epsilon_{m\bm{k}_0} \neq \epsilon_{\ell \bm{k}_0}$ ,
    \item[(ii)]  $\epsilon_{n\bm{k}_0} \neq \epsilon_{m\bm{k}_0}$ and $\epsilon_{m\bm{k}_0} = \epsilon_{\ell \bm{k}_0}$ ,
    \item[(iii)] $\epsilon_{n\bm{k}_0} = \epsilon_{m\bm{k}_0} = \epsilon_{\ell \bm{k}_0}$ .
\end{itemize}
Since $\mathcal{R}_{nm\ell} = \mathcal{R}_{mn\ell}$, it is sufficient to classify in the above manner.
In the case (i), the asymptotic form of $\mathcal{R}_{nm\ell}$ near $\bm{k}=\bm{k}_0$ is
\begin{align}
    \mathcal{R}_{nm\ell} &= \frac{-1}{\epsilon_{m\ell}^2} \left\{
    - \frac{2\int_{m\ell}}{\epsilon_{m\ell}} + f_m + f_{\ell} \right. \nonumber \\
    &+\left( \frac{3\int_{m\ell}}{\epsilon_{m\ell}^2}
    - \frac{2f_m + f_{\ell}}{\epsilon_{m\ell}} + \frac{1}{2} f'_m \right) \epsilon_{nm} \nonumber \\
    &+\left. \left( - \frac{4\int_{m\ell}}{\epsilon_{m\ell}^3} + \frac{3f_m + f_{\ell}}{\epsilon_{m\ell}^2}
    - \frac{f'_m}{\epsilon_{m\ell}} + \frac{1}{6} f''_m \right) \epsilon_{nm}^2 \right\} \nonumber \\
    &+ O(\epsilon_{nm}^3) .
    \label{eq:R_expand_i}
\end{align}
In the case (ii), the asymptotic form is
\begin{align}
    \mathcal{R}_{nm\ell} &=
    \frac{1}{\epsilon_{\ell n}} \left\{
    - \frac{\int_{\ell n}}{\epsilon_{\ell n}^2} + \frac{f_{\ell}}{\epsilon_{\ell n}} - \frac{1}{2} f'_{\ell} \right. \nonumber \\
    &+\left( \frac{\int_{\ell n}}{\epsilon_{\ell n}^3}
    - \frac{f_{\ell}}{\epsilon_{\ell n}^2} + \frac{f'_{\ell}}{2\epsilon_{\ell n}} -\frac{1}{6} f''_{\ell} \right) \epsilon_{m \ell} \nonumber \\
    &+\left. \left( - \frac{\int_{\ell n}}{\epsilon_{\ell n}^4} + \frac{f_{\ell}}{\epsilon_{\ell n}^3}
    - \frac{f'_{\ell}}{2\epsilon_{\ell n}^2} + \frac{f''_{\ell}}{6\epsilon_{\ell n}} - \frac{1}{24} f'''_{\ell} \right) \epsilon_{m \ell}^2 \right\} \nonumber \\
    &+ O(\epsilon_{m \ell}^3) .
    \label{eq:R_expand_ii}
\end{align}
In the case (iii), although $\mathcal{R}_{nm\ell} = -f''_n/6$ holds at $\bm{k}=\bm{k}_0$,
the asymptotic form depends on the order in which the limits are taken,
as evident from Eqs.~\eqref{eq:R_expand_i} and \eqref{eq:R_expand_ii}.

It has been shown that the integrand in Eq.~\eqref{eq:Mijk} does not diverge at the band touching point $\bm{k}_0$.
Equations~\eqref{eq:P_expand}-\eqref{eq:R_expand_ii} are useful for numerical calculations.

\subsection*{Model parameters}
We used the following hoppings in Eq.~\eqref{eq:Hk_for_Para},
which was obtained in Ref.~\cite{roig:prb2024}
as a minimal model of altermagnets for point group $D_{4h}$:
\begin{align}
    \varepsilon_{0,\bm{k}} &= \nonumber \\
    &\hspace{-1em}t_1 ( \cos k_x + \cos k_y ) - \mu + t_2 \cos k_z + t_3 \cos k_x \cos k_y \nonumber \\
    &\hspace{-1em}+t_4 ( \cos k_x + \cos k_y ) \cos k_z + t_5 \cos k_x \cos k_y \cos k_z , \\
    t_{x,\bm{k}}
    &= t_8 \cos \frac{k_x}{2} \cos \frac{k_y}{2} \cos \frac{k_z}{2} , \\
    t_{z,\bm{k}}
    &= t_6 \sin k_x \sin k_y + t_7 \sin k_x \sin k_y \cos k_z , \\
    \lambda_{x,\bm{k}}
    &= \lambda \sin \frac{k_z}{2} \sin \frac{k_x}{2} \cos \frac{k_y}{2} , \\
    \lambda_{y,\bm{k}}
    &= -\lambda \sin \frac{k_z}{2} \sin \frac{k_y}{2} \cos \frac{k_x}{2} , \\
    \lambda_{z,\bm{k}}
    &= \lambda_z \cos \frac{k_z}{2} \cos \frac{k_x}{2} \cos \frac{k_y}{2}
    (\cos k_x - \cos k_y) .
\end{align}
The hopping parameters are set to
$t_1=0$, $t_2=0.13$, $t_3=0$, $t_4=-0.02$, $t_5=0.015$,
$t_6=0$, $t_7=0.03$, $t_8=0.33$, and $\mu=-0.01$ in the main text
to reproduce the nonmagnetic band structures of MnF$_2$~\cite{roig:prb2024}.
We denote $\lambda_z$ as $\lambda$ throughout this paper, 
as we consider the cases where $\bm{\lambda}_{\bm{k}} = (\lambda_{x,\bm{k}}, 0, 0)$, 
$(0, 0, \lambda_{z,\bm{k}})$, or $(\lambda_{x,\bm{k}}, \lambda_{y,\bm{k}}, 0)$.

\textit{Note added.}
While preparing the manuscript for publication, we became aware of an independent parallel work%
~\cite{oike:prb2025spinMO} in which the expression for spin magnetic octupoles was also derived, in agreement with our findings.
Beyond this fundamental result, the two studies address complementary aspects of the problem; for example, this work focuses on their relevance to AFMs exhibiting the anomalous Hall effect. 

\section*{Data Availability}
The datasets generated and/or analyzed during the current study are not publicly available 
due to the absence of deposition in a public repository, but are available from the corresponding author on reasonable request.

\section*{Code Availability}
The numerical calculation codes used in this study are available from the corresponding author upon reasonable request.

\section*{Acknowledgments}
The authors gratefully acknowledge J. 
\ifmmode \bar{O}\else \={O}\fi{}ik\'e, K. Shinada, and R. Peters
for useful discussions and sharing a related unpublished manuscript prior to its submission.
S.H. was supported by JSPS KAKENHI Grants Numbers JP21H01037, JP22H00101, JP22H01183, JP23H04869, JP23K03288, JP23K20827, and by JST CREST (JPMJCR23O4) and JST FOREST (JPMJFR2366).

\section*{Author Contributions}
T.S. and S.H. conceived the project. 
T.S. performed the analytical and numerical calculations.
Both authors contributed to writing the paper.

\section*{Competing Interests}
The authors declare no competing financial or non-financial interests.

\section*{References}
%

\clearpage
\begin{table*}[htbp]
    \caption{
        \textbf{Classification of responses under magnetic multipole orderings beyond symmetry considerations.}
        Symmetry allowed responses under the magnetic multipole orderings are classified into three types
        based on the thermodynamic properties in insulators at $T=0$.
        Type~I (II) responses are characterized by the 
        first-order (higher-order) derivative of the magnetic multipole with respect to $\mu$,
        whereas Type~III responses do not appear in the thermodynamic arguments.
        For Type~I and Type~II responses unique to orbital magnetic multipoles, we denote ``(orbital)'' after the name of the response.
        The MQ and MO orders are characterized by the reducible tensors $M_{ij}$ and $M_{ijk}$, respectively.
        $M_{ij}$ is decomposed as $M_{ij} = M^{M_{2m}}_{ij} + M^{T_{1m}}_{ij} + M^{M_{0}}_{ij}$, 
        where each term represents, in order, the contributions from (symmetric) MQ, magnetic toroidal dipole, and magnetic monopole,
        satisfying the decomposition of a 9-dimensional reducible representation of SO(3): $9 = 5 \oplus 3 \oplus 1$.
        The decomposition of $M_{ijk}$, 
        which has the contributions from the (totally symmetric) MO, MTQ, and two MDs 
        is discussed around Eq.~\eqref{eq:MO_decomposition}.
    }
    \begin{ruledtabular}
    \renewcommand{\arraystretch}{1.3}
    \begin{tabular}{c|c|c|c}
        & 
        magnetic dipole $M_i$
        & 
        magnetic quadrupole $M_{ij}$
        & 
        magnetic octupole $M_{ijk}$
        \\
        \colrule
        (I)%
        &
        \begin{tabular}{c}
            anomalous Hall (orbital) \\ $J_i = \sigma_{ij} E_j$
        \end{tabular} 
        &
        \begin{tabular}{c}
            magnetoelectric \\ $M_i = \alpha_{ij} E_j$, $P_j = \alpha_{ij} B_i$ \\
            quadrupolar anomalous Hall (orbital) \\ $J_i = \sigma_{ijk} \partial_{r_j} E_k$
        \end{tabular} 
        &
        \begin{tabular}{c}
            quadrupolar magnetoelectric \\ $M_i = \alpha_{ijk} \partial_{r_j} E_k$, $P_{k} = - \alpha_{ijk} \partial_{r_j} B_{i}$ \\ 
            piezomagnetic \\ $Q_{jk} = \alpha_{ijk} B_i$ \\
            octupolar anomalous Hall (orbital) \\ $J_i = \sigma_{ijk\ell} \partial_{r_j} \partial_{r_k} E_{\ell}$
        \end{tabular} \\
        \colrule
        (II)%
        &
        &
        \begin{tabular}{c}
            second-order anomalous Hall (orbital) \\ $J_i = \sigma'_{ijk} E_j E_k$
        \end{tabular} 
        & 
        \begin{tabular}{c}
            second-order magnetoelectric \\ $M_i = \alpha'_{ijk} E_j E_k$ \\
            second-order quadrupolar anomalous Hall (orbital) \\ $J_i = \sigma'_{ijk\ell} (\partial_{r_j} E_k E_{\ell} + (j \leftrightarrow k \leftrightarrow \ell))$ \\
            third-order anomalous Hall (orbital) \\ $J_i = \sigma''_{ijk\ell} E_j E_k E_{\ell}$
        \end{tabular}\\
        \colrule
        (III)%
        & 
        \begin{tabular}{c}
            e.g., spin current generation \\ (dissipative) \\ $J_i \sigma_j = \sigma^{\mathrm{S}}_{ijk} J_k$
        \end{tabular} & 
        \begin{tabular}{c}
            e.g., current-induced distortion \\ (dissipative) \\ $\zeta_{ij} = d_{ijk} J_k$
        \end{tabular} &
        \begin{tabular}{c}
            e.g., anomalous Hall \\ $J_i = \sigma_{ij} E_j$ \\
            spin current generation (dissipative) \\ $J_i \sigma_j = \sigma^{\mathrm{S}}_{ijk} J_k$
        \end{tabular}
    \end{tabular}
\end{ruledtabular}
    \label{tab:response}
\end{table*}

\begin{table*}[htbp]
    \caption{
        \textbf{Relations among TRSB terms and various quantities.}
        Summary of the TRSB term in Eqs.~\eqref{eq:Hk_for_AFM} and \eqref{eq:Hk_for_FM},
        nonzero component of the symmetry-allowed axial vector $\bm{h}=(h_x, h_y, h_z)$,
        dominant MD,
        and nonzero component of the Hall conductivity tensor $\sigma_{ij}$.
    }
    \begin{ruledtabular}
    \renewcommand{\arraystretch}{1.3}
    \begin{tabular}{cccccccc}
        & & AFM (Eq.~\eqref{eq:Hk_for_AFM}) & & \quad & & FM (Eq.~\eqref{eq:Hk_for_FM}) & \\
        TRSB term & $\hat{\tau}_z \hat{\sigma}_x$ & $\hat{\tau}_z \hat{\sigma}_y$
        & $\hat{\tau}_z \hat{\sigma}_z$ & \quad & $\hat{\sigma}_x$
        & $\hat{\sigma}_y$ & $\hat{\sigma}_z$ \\
        \colrule
        $\bm{h}$ & $h_y$ & $h_x$ & None & \quad & $h_x$ & $h_y$ & $h_z$ \\
        \colrule
        Dominant MD & $M'_y$ & $M'_x$ & None & \quad & $M_x$ & $M_y$ & $M_z$ \\
        \colrule
        $\sigma_{ij}$ & $\sigma_{zx}$ & $\sigma_{yz}$ & None & \quad & $\sigma_{yz}$ & $\sigma_{zx}$ & $\sigma_{xy}$
    \end{tabular}
\end{ruledtabular}
    \label{tab:axial}
\end{table*}

\clearpage
\onecolumngrid
\begin{center}
\textbf{\large Supplemental Material for\\
``Quantum theory of magnetic octupole in periodic crystals and 
application to $d$-wave altermagnets''}\\
\vspace{1em}
{Takumi Sato and Satoru Hayami}\\
\textit{Graduate School of Science, Hokkaido University, Sapporo 060-0810, Japan}\\
\end{center}
\vspace{0.5cm}

\setcounter{section}{0}
\setcounter{equation}{0}
\setcounter{figure}{0}
\setcounter{table}{0}
\renewcommand{\thesection}{S\arabic{section}}
\renewcommand{\theequation}{S\arabic{equation}}
\renewcommand{\thefigure}{S\arabic{figure}}
\renewcommand{\thetable}{S\arabic{table}}

\section{Tensor decomposition of time-reversal-odd rank-3 axial tensor}
\label{sec:tensor_decomposition}
The time-reversal-odd rank-3 axial tensor $M_{ijk}$ is decomposed as
\begin{align}
    M_{ijk} &= M_{ijk}^{M_{3m}} + M_{ijk}^{T_{2m}} + M_{ijk}^{M_{1m}} + M_{ijk}^{M'_{1m}} ,
\end{align}
where $M_{ijk}^{M_{3m}}$, $M_{ijk}^{T_{2m}}$, $M_{ijk}^{M_{1m}}$, and $M_{ijk}^{M'_{1m}}$ represent
the component of (totally symmetric) magnetic octupole (MO), magnetic toroidal quadrupole (MTQ), 
magnetic dipole (MD), and anisotropic magnetic dipole (AMD), respectively%
~\cite{urru:annphys2022,kusunose:jpsj2020}.
The components of MO in $M_{ijk}$ are
\begin{align}
    M_{xjk}^{M_{3m}} &= \left( \begin{array}{ccc}
        \frac{1}{5} M_{x}^{\alpha} & -\frac{1}{10}M_{y}^{\alpha}-\frac{1}{2}M_{y}^{\beta} & -\frac{1}{10}M_{z}^{\alpha}+\frac{1}{2}M_{z}^{\beta} \\
        -\frac{1}{10}M_{y}^{\alpha}-\frac{1}{2}M_{y}^{\beta} & -\frac{1}{10}M_{x}^{\alpha}+\frac{1}{2}M_{x}^{\beta} & M_{[xyz]} \\
        -\frac{1}{10}M_{z}^{\alpha}+\frac{1}{2}M_{z}^{\beta} & M_{[xyz]} & -\frac{1}{10}M_{x}^{\alpha}-\frac{1}{2}M_{x}^{\beta}
    \end{array} \right) 
    \label{eq:Mxjk_MO} , \\
    M_{yjk}^{M_{3m}} &= \left( \begin{array}{ccc}
        -\frac{1}{10}M_{y}^{\alpha}-\frac{1}{2}M_{y}^{\beta} & -\frac{1}{10}M_{x}^{\alpha}+\frac{1}{2}M_{x}^{\beta} & M_{[xyz]} \\
        -\frac{1}{10}M_{x}^{\alpha}+\frac{1}{2}M_{x}^{\beta} & \frac{1}{5}M_{y}^{\alpha} & -\frac{1}{10}M_{z}^{\alpha}-\frac{1}{2}M_{z}^{\beta} \\
        M_{[xyz]} & -\frac{1}{10}M_{z}^{\alpha}-\frac{1}{2}M_{z}^{\beta} & -\frac{1}{10}M_{y}^{\alpha}+\frac{1}{2}M_{y}^{\beta}
    \end{array} \right)
    \label{eq:Myjk_MO} , \\
    M_{zjk}^{M_{3m}} &= \left( \begin{array}{ccc}
        -\frac{1}{10}M_{z}^{\alpha}+\frac{1}{2}M_{z}^{\beta} & M_{[xyz]} & -\frac{1}{10}M_{x}^{\alpha}-\frac{1}{2}M_{x}^{\beta} \\
        M_{[xyz]} & -\frac{1}{10}M_{z}^{\alpha}-\frac{1}{2}M_{z}^{\beta} & -\frac{1}{10}M_{y}^{\alpha}+\frac{1}{2}M_{y}^{\beta} \\
        -\frac{1}{10}M_{x}^{\alpha}-\frac{1}{2}M_{x}^{\beta} & -\frac{1}{10}M_{y}^{\alpha}+\frac{1}{2}M_{y}^{\beta} & \frac{1}{5}M_{z}^{\alpha}
    \end{array} \right)
    \label{eq:Mzjk_MO} ,
\end{align}
where
\begin{align}
    M_{[xyz]}      &\coloneqq \frac{1}{3} ( M_{xyz} + M_{zxy} + M_{yzx} ) , \\
    M_{x}^{\alpha} &\coloneqq 2M_{xxx} - ( M_{xyy} + M_{xzz} ) - 2( M_{yyx} + M_{zzx} ) , \\
    M_{y}^{\alpha} &\coloneqq 2M_{yyy} - ( M_{yxx} + M_{yzz} ) - 2( M_{xxy} + M_{zzy} ) , \\
    M_{z}^{\alpha} &\coloneqq 2M_{zzz} - ( M_{zxx} + M_{zyy} ) - 2( M_{xxz} + M_{yyz} ) , \\
    M_{x}^{\beta}  &\coloneqq \frac{1}{3} \left\{ M_{xyy} - M_{xzz} + 2( M_{yyx} - M_{zzx} ) \right\} , \\
    M_{y}^{\beta}  &\coloneqq \frac{1}{3} \left\{ M_{yzz} - M_{yxx} + 2( M_{zzy} - M_{xxy} ) \right\} , \\
    M_{z}^{\beta}  &\coloneqq \frac{1}{3} \left\{ M_{zxx} - M_{zyy} + 2( M_{xxz} - M_{yyz} ) \right\} ,
\end{align}
are MOs.
The components of MTQ in $M_{ijk}$ are
\begin{align}
    M_{xjk}^{T_{2m}} &= \frac{1}{3} \left( \begin{array}{ccc}
        0 & -T_{zx} & T_{xy} \\
        -T_{zx} & -2T_{yz} & -(T_{v} + 3T_{u}) \\
        T_{xy} & -(T_{v} + 3T_{u}) & 2T_{yz}
    \end{array} \right)
    \label{eq:Mxjk_MTQ} , \\
    M_{yjk}^{T_{2m}} &= \frac{1}{3} \left( \begin{array}{ccc}
        2T_{zx} & T_{yz} & -(T_{v} - 3T_{u}) \\
        T_{yz} & 0 & -T_{xy} \\
        -(T_{v} - 3T_{u}) & -T_{xy} & -2T_{zx}
    \end{array} \right)
    \label{eq:Myjk_MTQ} , \\
    M_{zjk}^{T_{2m}} &= \frac{1}{3} \left( \begin{array}{ccc}
        -2T_{xy} & 2T_{v} & -T_{yz} \\
        2T_{v} & 2T_{xy} & T_{zx} \\
        -T_{yz} & T_{zx} & 0
    \end{array} \right)
    \label{eq:Mzjk_MTQ} ,
\end{align}
where
\begin{align}
    T_{u} &\coloneqq -\frac{1}{2} \epsilon_{ijz} M_{ijz} , \\
    T_{v} &\coloneqq -\frac{1}{2} ( \epsilon_{ijx} M_{ijx} - \epsilon_{ijy} M_{ijy} ) , \\
    T_{yz} &\coloneqq -\frac{1}{2} ( \epsilon_{ijz} M_{ijy} + \epsilon_{ijy} M_{ijz} ) , \\
    T_{zx} &\coloneqq -\frac{1}{2} ( \epsilon_{ijx} M_{ijz} + \epsilon_{ijz} M_{ijx} ) , \\
    T_{xy} &\coloneqq -\frac{1}{2} ( \epsilon_{ijy} M_{ijx} + \epsilon_{ijx} M_{ijy} ) ,
\end{align}
with $\epsilon_{ijk}$ being the totally antisymmetric tensor,
are MTQs.
The components of MD in $M_{ijk}$ are
\begin{align}
    M_{xjk}^{M_{1m}} = \frac{1}{3} \left( \begin{array}{ccc}
        M_{x} & 0 & 0 \\
        0 & M_{x} & 0 \\
        0 & 0 & M_{x}
    \end{array} \right)
    \label{eq:Mxjk_MD} , \\
    M_{yjk}^{M_{1m}} = \frac{1}{3} \left( \begin{array}{ccc}
        M_{y} & 0 & 0 \\
        0 & M_{y} & 0 \\
        0 & 0 & M_{y}
    \end{array} \right)
    \label{eq:Myjk_MD} , \\
    M_{zjk}^{M_{1m}} = \frac{1}{3} \left( \begin{array}{ccc}
        M_{z} & 0 & 0 \\
        0 & M_{z} & 0 \\
        0 & 0 & M_{z}
    \end{array} \right)
    \label{eq:Mzjk_MD} ,
\end{align}
where
\begin{align}
    M_{i} &\coloneqq M_{ijj} ,
\end{align}
is an MD.
The components of AMD in $M_{ijk}$ are
\begin{align}
    M_{xjk}^{M'_{1m}} &= \frac{\sqrt{10}}{15} \left( \begin{array}{ccc}
        2M'_{x} & \frac{3}{2}M'_{y} & \frac{3}{2}M'_{z} \\
        \frac{3}{2}M'_{y} & -M'_{x} & 0 \\
        \frac{3}{2}M'_{z} & 0 & -M'_{x}
    \end{array} \right)
    \label{eq:Mxjk_AMD} , \\
    M_{yjk}^{M'_{1m}} &= \frac{\sqrt{10}}{15} \left( \begin{array}{ccc}
        -M'_{y} & \frac{3}{2}M'_{x} & 0 \\
        \frac{3}{2}M'_{x} & 2M'_{y} & \frac{3}{2}M'_{z} \\
        0 & \frac{3}{2}M'_{z} & -M'_{y}
    \end{array} \right)
    \label{eq:Myjk_AMD} , \\
    M_{zjk}^{M'_{1m}} &= \frac{\sqrt{10}}{15} \left( \begin{array}{ccc}
        -M'_{z} & 0 & \frac{3}{2}M'_{x} \\
        0 & -M'_{z} & \frac{3}{2}M'_{y} \\
        \frac{3}{2}M'_{x} & \frac{3}{2}M'_{y} & 2M'_{z}
    \end{array} \right)
    \label{eq:Mzjk_AMD} ,
\end{align}
where
\begin{align}
    M'_{i} &\coloneqq \frac{1}{\sqrt{10}} ( 3M_{jji} - M_{ijj} ) ,
\end{align}
is an AMD.

\section{Effects of SOC in the collinear AFM}
\label{sec:comparison_SOC}
\begin{figure}[htbp]
    \includegraphics[width=8.0cm]{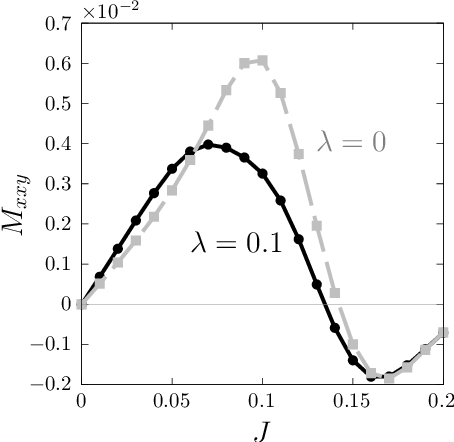}
    \caption{
            $J$
            dependence of $M_{xxy}$ in the collinear antiferromagnetic state of the MnF$_2$ model with $\bm{J}=(J,0,0)$.
            All parameters are the same as in Fig.~2(a) of the main text,
            except for the amplitude of the SOC.
    }
    \label{fig:Mijk_jdep_woSOC}
\end{figure}

Figure~\ref{fig:Mijk_jdep_woSOC} shows the $J$ 
dependence of $M_{xxy}$ 
in the collinear antiferromagnetic state of the MnF$_2$ model, 
comparing the cases with $\lambda = 0.1$ and $\lambda = 0$ (i.e., the vanishing SOC limit).
The other parameters are identical to those in Fig.~2(a) of the main text.
Both cases exhibit qualitatively similar behavior, 
indicating that SOC parallel to the magnetic moment has little effect on the magnetic structure.

\section{Origin of the sign change of the MO}
\label{sec:sign_change}
\begin{figure}[htbp]
    \includegraphics[width=8.0cm]{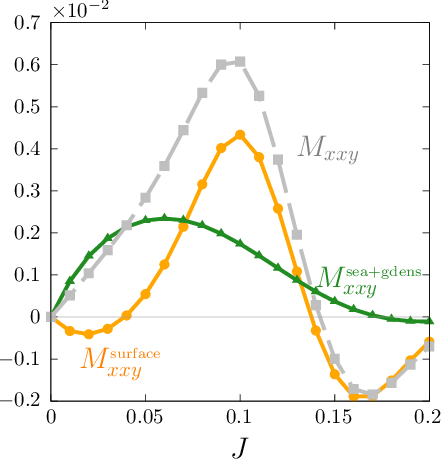}
    \caption{
            $J$
            dependence of $M_{xxy}$, $M^{\mathrm{surface}}_{xxy}$, and
            $M^{\mathrm{sea}+\mathrm{gdens}}_{xxy}$
            in the collinear antiferromagnetic state of the MnF$_2$ model with $\bm{J}=(J,0,0)$
            in the vanishing SOC limit.
            The other parameters are the same as those in Fig.~2(a) of the main text.
    }
    \label{fig:Mijk_jdep_FSurfaceTerm}
\end{figure}

In this section, we discuss the origin of the sign change of the MO.
As a specific example, we consider the collinear antiferromagnetic state of the MnF$_2$ model 
with $\bm{J}=(J,0,0)$ in the vanishing SOC limit, for simplicity.
As discussed in Sec.~\ref{sec:comparison_SOC} and the main text,
the SOC plays only a minor role in the collinear magnetic structure.
In this setup, only the spin-diagonal component $S_{n,\text{diag}}^{ijk}$ remains finite
in Eq.~(11) of the main text, while the spin-off-diagonal component
$S_{n,\text{off-diag}}^{ijk}$ vanishes.
We decompose $M_{ijk}$ into the Fermi-surface term and the remaining term as follows:
\begin{align}
    M_{ijk} &= M^{\mathrm{surface}}_{ijk} + M^{\mathrm{sea}+\mathrm{gdens}}_{ijk} ,
\end{align}
where
\begin{align}
    M^{\mathrm{surface}}_{ijk} &\coloneqq
    g \mu_{\mathrm{B}} \sum_{n} \int \frac{d^d k}{(2\pi)^d}
    \left( -\frac{1}{12} s^i_n \partial_{k_j} \partial_{k_k} \epsilon_n f_n' \right) , \\
    M^{\mathrm{sea}+\mathrm{gdens}}_{ijk} &\coloneqq
    g \mu_{\mathrm{B}} \sum_{n} \int \frac{d^d k}{(2\pi)^d}
    \sum_{m}^{\neq n} \left[
        \frac{1}{2} (s^i_n+s^i_m) \left\{ g^{jk}_{nm} f_n + G^{jk}_{nm} \int_{\epsilon_n-\mu}^{\infty} dz f(z) \right\}
    \right] .
\end{align}
We set $T=0.01$, as in the main text, 
and show the $J$ dependence of each term in Fig.~\ref{fig:Mijk_jdep_FSurfaceTerm}.
We find that the sign change of $M_{xxy}$ around $J \sim 0.14$ 
is mainly caused by $M^{\mathrm{surface}}_{xxy}$.
Therefore, we focus on analyzing the Fermi surfaces in what follows.

\begin{figure}[htbp]
    \includegraphics[width=16.0cm]{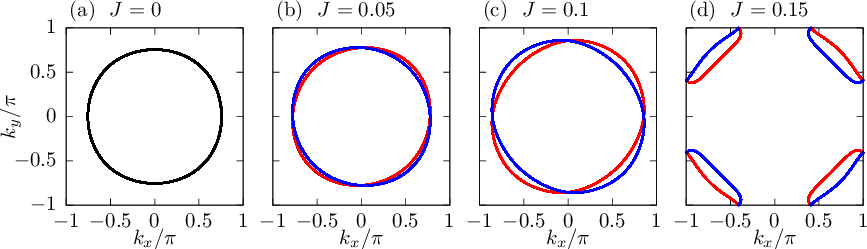}
    \caption{
            Fermi surfaces at $k_z=0$ in the collinear antiferromagnetic state 
            of the MnF$_2$ model without SOC, shown for
            several values of $J$.
            Red (blue) points indicate the Fermi surfaces 
            of the spin $\uparrow$ ($\downarrow$) bands.
    }
    \label{fig:FS_jdep}
\end{figure}

To understand the origin of the sign change of $M^{\mathrm{surface}}_{xxy}$ around $J \sim 0.14$, 
we show the Fermi surfaces at $k_z=0$ for several values of $J$ in Fig.~\ref{fig:FS_jdep}.
We find that the shape of the Fermi surface differs significantly
between $J=0.1$ [Fig.~\ref{fig:FS_jdep}(c)] and $J=0.15$ [(d)].
This difference appears to be the main reason for the sign change in $M^{\mathrm{surface}}_{xxy}$.
Although other factors such as temperature 
or the shape of the Fermi surface in other $k_z$-planes 
may also influence $M^{\mathrm{surface}}_{xxy}$,
we believe that the deformation of the Fermi surface captured in Fig.~\ref{fig:FS_jdep}
plays an important role.

\clearpage
\section{Band structures under a different parameter setting}
\label{sec:band_RuO2}
\begin{figure}[htbp]
    \includegraphics[width=8.0cm]{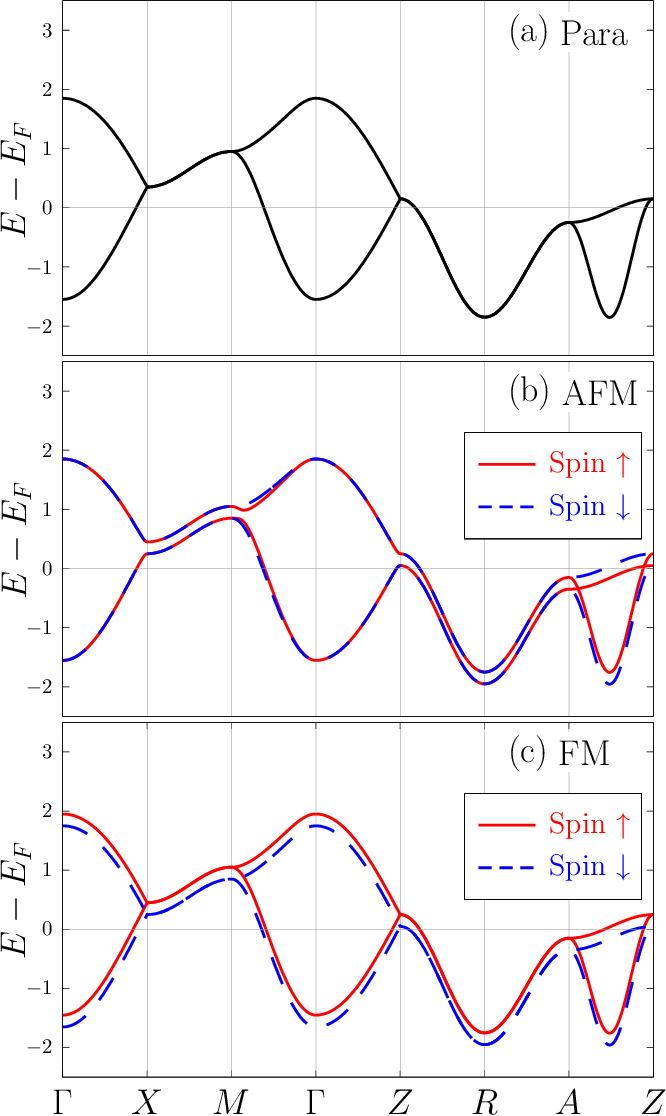}
    \caption{
        Band structures in the paramagnetic (a), the antiferromagnetic (b), and
        the ferromagnetic states (c) without SOC.
        The parameters are chosen to reproduce
        the paramagnetic band structures of RuO$_2$ from $d_{xy}$ orbital, which is obtained
        in Ref.~\cite{roig:prb2024}
        and $J=0.1$ in (b) and (c).
    }
    \label{fig:band_RuO2}
\end{figure}

We present numerical results for the MO
obtained by using a different set of model parameters in $\hat{H}_{\bm{k}}^{\mathrm{Para}}$.
Here, we choose the parameters
for the one-orbital model of RuO$_2$ listed in Ref.~\cite{roig:prb2024},
where the point group symmetry in the Hamiltonian without SOC is $D_{4h}$.
The hopping parameters are $t_1=-0.05$, $t_2=0.7$, $t_3=0.5$, $t_4=-0.15$, $t_5=-0.4$,
$t_6=-0.6$, $t_7=0.3$, $t_8=1.7$, and $\mu=0.25$.
The band structures without the SOC (i.e., $\lambda=0$)
in the paramagnetic state, the antiferromagnetic state, and the ferromagnetic state
are shown in Fig.~\ref{fig:band_RuO2}(a), (b), and (c), respectively.
In (b) and (c), we choose $J=0.1$ as in the main text.
The spin splitting along the $M-\Gamma$ and $A-Z$ lines is obtained,
and hence this model in the antifferomagnetic state also describes
the electronic structures of the $d_{xy}$-wave altermagnet.

\section{Numerical results of the MO for the different parameter setting}
\label{sec:results_RuO2}
\begin{figure}[htbp]
    \includegraphics[width=8.5cm]{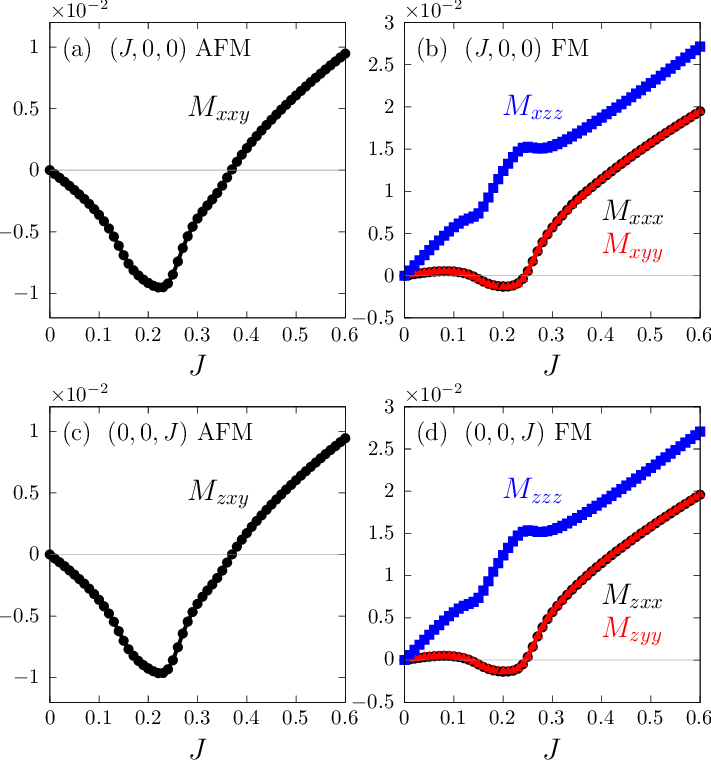}
    \caption{
        [(a) and (b)] ([(c) and (d)])
        $J$ dependence of the
        nonzero components of $M_{ijk}$ for $\bm{J}=(J,0,0)$ ($\bm{J}=(0,0,J)$) in the AFM and FM, respectively.
        We choose $\lambda=0.1$ and $T=0.01$.
        The parameters for the one-orbital model of RuO$_2$ are used.
    }
    \label{fig:Mijk_jdep_RuO2}
\end{figure}

In Fig.~\ref{fig:Mijk_jdep_RuO2}, the $J$ dependence of the MO for the above parameter set is presented.
In (a) and (b) [(c) and (d)], we show the nonzero components of the MO for $\bm{J}=(J,0,0)$ [$\bm{J}=(0,0,J)$]
in the AFM and FM, respectively.
As in the main text, we assumed that the SOC has the same orientation as $\bm{J}$.
We set $\lambda = 0.1$ and $T=0.01$.
Although the $J$ dependence seems to differ from that for MnF$_2$ in the main text,
both models exhibit an almost linear dependence on $J$ when $J$ is small compared with the bandwidth.
When $J$ becomes sufficiently large, the MO similarly shows nonmonotonic behavior and changes its sign.

\begin{figure}[htbp]
    \includegraphics[width=8.0cm]{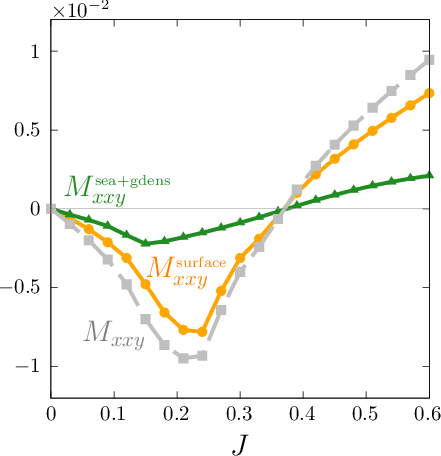}
    \caption{
        $J$
        dependence of $M_{xxy}$, $M^{\mathrm{surface}}_{xxy}$, and
        $M^{\mathrm{sea}+\mathrm{gdens}}_{xxy}$
        in the collinear antiferromagnetic state of the RuO$_2$ model with $\bm{J}=(J,0,0)$
        in the vanishing SOC limit.
        The other parameters are the same as those in Fig.~S5(a).
    }
    \label{fig:Mijk_jdep_FSurfaceTerm_RuO2}
\end{figure}

Next, to investigate the origin of the sign change, 
we discuss the properties of the decomposed MO, as in Sec.~\ref{sec:sign_change},
for the RuO$_2$ model in the vanishing SOC limit ($\lambda=0$) with $\bm{J}=(J,0,0)$.
In Fig.~\ref{fig:Mijk_jdep_FSurfaceTerm_RuO2}, we show the $J$ dependence of 
$M_{xxy}$, $M^{\mathrm{surface}}_{xxy}$, and $M^{\mathrm{sea}+\mathrm{gdens}}_{xxy}$.
The other parameters are the same as those in Fig.~\ref{fig:Mijk_jdep_RuO2}(a).
We find that the sign change of $M_{xxy}$ around $J \sim 0.39$ is mainly caused by $M^{\mathrm{surface}}_{xxy}$.
Therefore, we next discuss the shape of the Fermi surface.

\begin{figure}[htbp]
    \includegraphics[width=16.0cm]{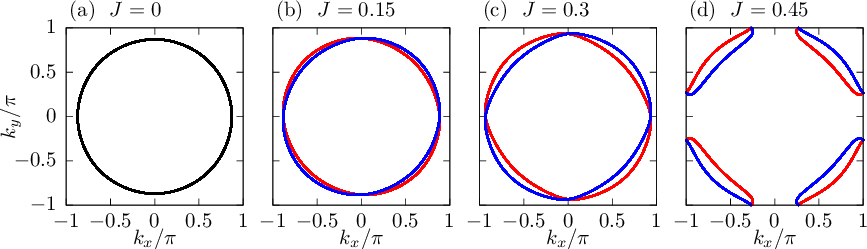}
    \caption{
        Fermi surfaces at $k_z=0$ in the collinear antiferromagnetic state 
        of the RuO$_2$ model without SOC, shown for
        several values of $J$.
        Red (blue) points indicate the Fermi surfaces 
        of the spin $\uparrow$ ($\downarrow$) bands.
    }
    \label{fig:FS_jdep_RuO2}
\end{figure}

In Fig.~\ref{fig:FS_jdep_RuO2}, we show the Fermi surfaces at $k_z=0$ for several values of $J$.
We find that the shape of the Fermi surface differs significantly
between $J=0.3$ [Fig.~\ref{fig:FS_jdep_RuO2}(c)] and $J=0.45$ [(d)].
This difference appears to be the main reason for the sign change 
in $M^{\mathrm{surface}}_{xxy}$, similar to the MnF$_2$ model.
We believe that the deformation of the Fermi surface plays a dominant role even in this model.

\begin{figure}[htbp]
    \includegraphics[width=8.5cm]{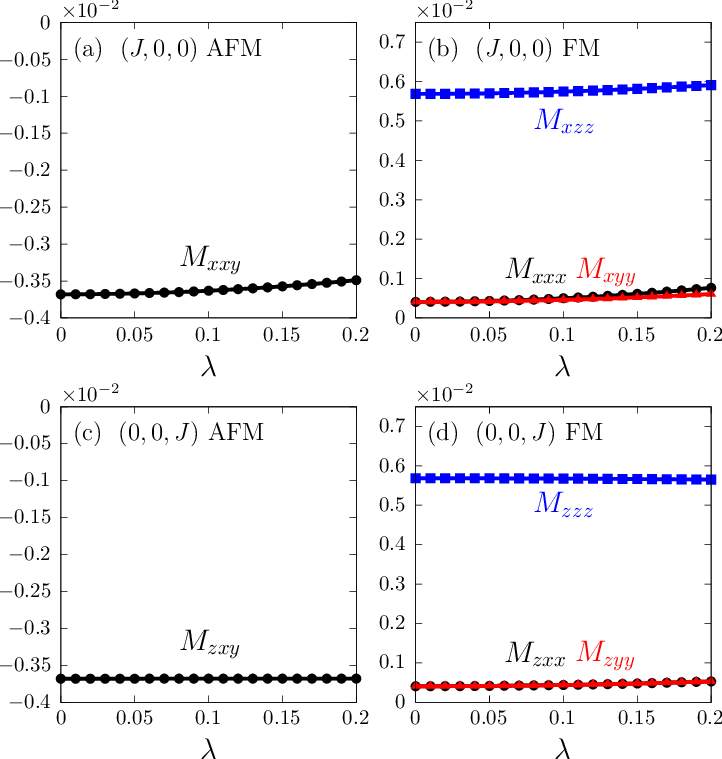}
    \caption{
        [(a) and (b)] ([(c) and (d)])
        $\lambda$ dependence of the
        nonzero components of $M_{ijk}$ when $\bm{J}=(J,0,0)$ ($\bm{J}=(0,0,J)$) in the AFM and FM, respectively.
        We choose $J=0.1$ and $T=0.01$.
        The parameters for the one-orbital model of RuO$_2$ are used.
    }
    \label{fig:Mijk_lamdep_RuO2}
\end{figure}

In Fig.~\ref{fig:Mijk_lamdep_RuO2},
the $\lambda$ dependence of the nonzero components of the MO is shown.
As in the main text, we choose $\bm{\lambda}_{\bm{k}} \parallel \bm{J}$, $J=0.1$, and $T=0.01$.
The MO is almost independent of the SOC, and hence the SOC has the little effect on the magnetic structure.
This feature is the same as that for MnF$_2$ discussed in the main text.

\begin{figure}[htbp]
    \includegraphics[width=8.5cm]{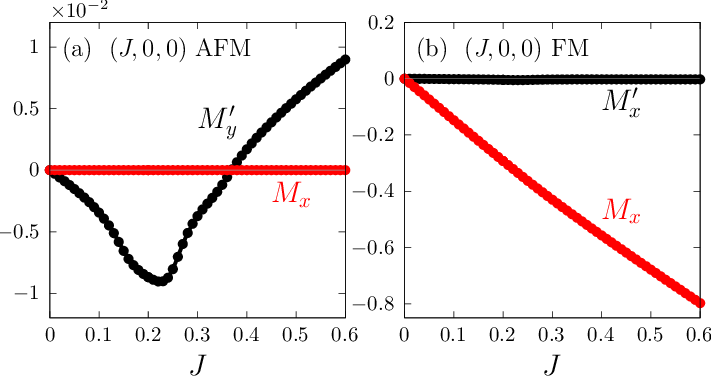}
    \caption{
        $J$ dependence of the finite components of the AMD ($M'_{i}$) and the spin magnetization ($M_{i}$)
        in the AFM (FM) is shown in (a) ((b)), where the vertical axis is expressed in units of $\mu_{\mathrm{B}}$.
        Although the spin magnetization remains zero in the AFM, it is shown for reference.
        The parameters are the same as those in Fig.~\ref{fig:Mijk_jdep_RuO2}.
    }
    \label{fig:AMD_RuO2}
\end{figure}

In Fig.~\ref{fig:AMD_RuO2}, the nonzero components of the AMD and the spin magnetization are shown
as a function of $J$.
We adopt the same situation as the main text:
$\bm{J}=(J,0,0)$, $\bm{\lambda}_{\bm{k}} \parallel \bm{J}$, $\lambda=0.1$, and $T=0.01$.
In the AFM (FM), $M'_y=3M_{xxy}/\sqrt{10}$ ($M_x$) increases almost linearly with increasing $J$
when $J$ is small.
Therefore, the MD that characterizes the magnetic structure of the AFM (FM)
with $\bm{J}=(J,0,0)$ is identified as $M'_y$ ($M_x$).

Although the detailed behavior in Figs.~\ref{fig:Mijk_jdep_RuO2}-\ref{fig:AMD_RuO2}
differs from that 
of the MnF$_2$ model, the overall trend for the one-orbital model of RuO$_2$ remains almost the same.
This supports our conclusions in the main text.

\section{Comparison between the AHC and the MO}
\label{sec:AHC}
\begin{figure}[htbp]
    \includegraphics[width=8.5cm]{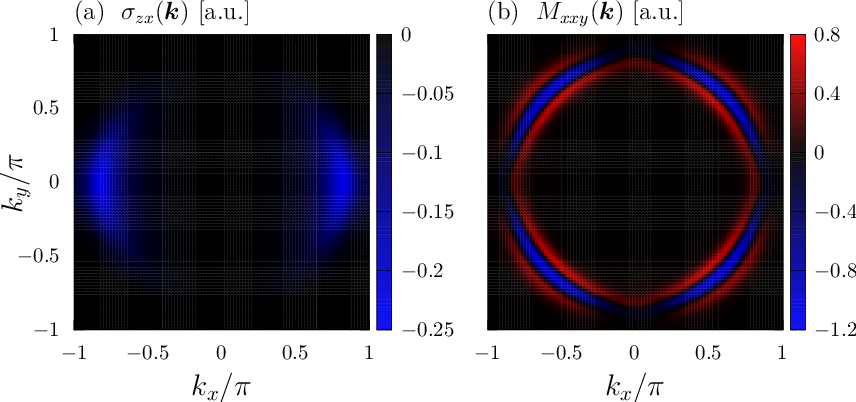}
    \caption{
        $k$-resolved AHC $\sigma_{zx}(\bm{k})$ and MO $M_{xxy}(\bm{k})$ at $k_z = 0$ 
        in the first Brillouin zone are shown in (a) and (b), respectively. 
        The antiferromagnetic state of the MnF$_2$ model with $\bm{J} = (J, 0, 0)$ and $\bm{\lambda}_{\bm{k}} \parallel \bm{J}$ 
        is chosen for the calculation. 
        The parameters are set as follows: $J = 0.1$, $\lambda = 0.1$, and $T = 0.01$.
    }
    \label{fig:AHC_MO_kdist}
\end{figure}

In Fig.~\ref{fig:AHC_MO_kdist}, we show the momentum-space distributions 
of the anomalous Hall conductivity (AHC) $\sigma_{ij}$ and the MO $M_{ijk}$ at $k_z=0$.
The calculation is performed for the collinear antiferromagnetic MnF$_2$ model 
with $\bm{J} = (J, 0, 0)$ and $\bm{\lambda}_{\bm{k}} \parallel \bm{J}$, 
using parameters $J = 0.1$, $\lambda = 0.1$, and $T = 0.01$.
In this setting, both $\sigma_{zx}$ and $M_{xxy}$ take finite values.
The $k$-resolved quantities are defined as
\begin{align}
    \sigma_{zx} &= \int \frac{d^3k}{(2\pi)^3} \sigma_{zx}(\bm{k}), \quad
    M_{xxy} = \int \frac{d^3k}{(2\pi)^3} M_{xxy}(\bm{k}) .
\end{align}
From Fig.~\ref{fig:AHC_MO_kdist}, we do not find a clear correlation between the AHC and the MO.
While $M_{xxy}(\bm{k})$ tends to be enhanced near the Fermi surface, $\sigma_{zx}(\bm{k})$ shows no such tendency.
Moreover, the SOC does not affect $M_{xxy}(\bm{k})$ on the $k_z=0$ plane,
since $\lambda_{x,\bm{k}} \propto \lambda \sin (k_z/2)$.
In contrast, $\sigma_{zx}(\bm{k})$ is influenced by the SOC.
These observations suggest that the relationship between the AHC and the MO is indirect.

\section{Effects of temperature}
\label{sec:temp_dep}
\begin{figure}[htbp]
    \includegraphics[width=8.0cm]{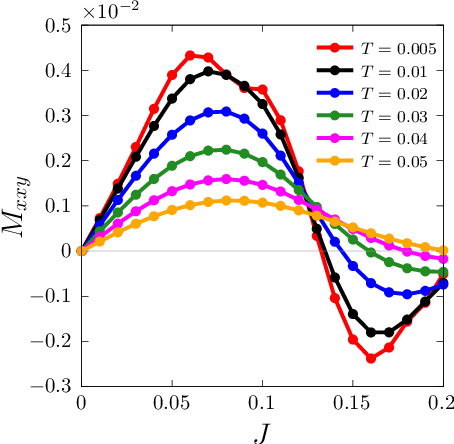}
    \caption{
        $J$ dependence of $M_{xxy}$ in the collinear antiferromagnetic state 
        of the MnF$_2$ model with $\bm{J}=(J,0,0)$,
        calculated at varying temperatures.
        The other parameters are the same as those in Fig.~2(a) of the main text;
        therefore, the black line-and-symbol plot corresponds to that in Fig.~2(a).
    }
    \label{fig:temp}
\end{figure}

In Fig.~\ref{fig:temp}, we show the $\bm{J}$ dependence 
of the finite component of the MO in the collinear antiferromagnetic state 
of the MnF$_2$ model, where the temperature is varied from $T=0.005$ to $T=0.05$.
The other parameters are identical to those in Fig.~2(a) of the main text.
As the temperature increases, the magnitude of the MO tends to decrease.
Moreover, the value of $J$ at which the sign change occurs also varies, 
implying the significance of the Fermi-surface term in this model.

\section{Induced components of the MO due to SOC}
\label{sec:noncollinear}
\begin{figure}[htbp]
    \includegraphics[width=8.5cm]{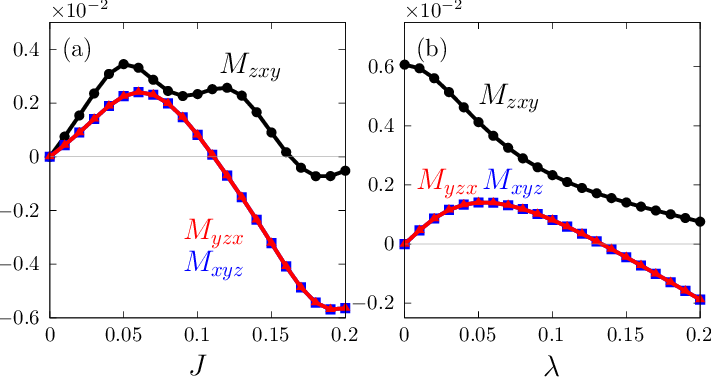}
    \caption{
        $J$ ($\lambda$) dependence of the finite components of the MO
        in the antiferromagnetic state of the MnF$_2$ model with 
        $\bm{\lambda}_{\bm{k}} = (\lambda_{x,\bm{k}}, \lambda_{y,\bm{k}}, 0)$ 
        and $\bm{J} = (0, 0, J)$ is shown in (a) ((b)).
        $\lambda = 0.1 / \sqrt{2}$ in (a) and $J=0.1$ in (b).
        The other parameters are the same as those in Fig.~2(c) of the main text.
    }
    \label{fig:noncollinear}
\end{figure}

In this section, we discuss the effect of noncollinear spin structures induced by the SOC.
As a representative and simple example, we consider the antiferromagnetic state of the MnF$_2$ model.
We assume $\bm{\lambda}_{\bm{k}} = (\lambda_{x,\bm{k}}, \lambda_{y,\bm{k}}, 0)$ 
and $\bm{J} = (0, 0, J)$, while all other parameters 
are the same as those in Fig.~2(c) of the main text.
Figure~\ref{fig:noncollinear}(a) ((b)) shows the $J$ ($\lambda$) 
dependence of the finite components of $M_{ijk}$ for $\lambda = 0.1 / \sqrt{2}$ ($J = 0.1$).
When SOC is absent or aligned along $(0, 0, J)$, only $M_{zxy}$ is finite.
Once the SOC is turned on, the additional components $M_{xyz} = M_{yzx}$ become finite.
These components contribute to MO 
$M_{[xyz]} = \frac{1}{3} ( M_{xyz} + M_{zxy} + M_{yzx} )$
and MTQ $T_{v} = M_{zxy}-M_{xyz}$.
The induced components become comparable to $M_{zxy}$ and exhibit nonmonotonic behavior
when $\lambda$ is sufficiently large.
Moreover, they change sign around $J \sim 0.11$ ($\lambda \sim 0.13$). 
We expect that the variation of the induced components
can be captured by analyzing the tensor components of 
the responses discussed in Results section of the main text.
We believe that our MO appropriately reflects the changes in the magnetic structure induced by the SOC. 
This issue will be investigated further.

\end{document}